\documentclass[letterpaper,twocolumn,10pt]{article}
\usepackage{usenix-2020-09}

\usepackage{tikz}
\usetikzlibrary{arrows,shadows,calc,matrix,positioning,tikzmark,patterns}

\definecolor{lblue}{HTML}{6E9CC4}
\definecolor{blue}{HTML}{41739F}
\definecolor{dblue}{HTML}{2F5474}
\definecolor{lyellow}{HTML}{FFE299}
\definecolor{yellow}{HTML}{FFBA08}
\definecolor{dyellow}{HTML}{B88400}
\definecolor{lred}{HTML}{F5B7BF}
\definecolor{red}{HTML}{bd0d00}
\definecolor{dred}{HTML}{bd0d00}
\definecolor{lgreen}{HTML}{94D388}
\definecolor{green}{HTML}{7AC86A}
\definecolor{dgreen}{HTML}{4EA43D}

\usepackage[utf8]{inputenc}\DeclareUnicodeCharacter{2212}{-} % this is needed for PGF files?

\newlength\unitlen
\newlength\unitlenbox
\newlength\unitlenboxheight
\newlength\unitlenboxwidth

\usepackage[ruled,vlined]{algorithm2e}

\newcommand{\uldlsniffer}{\textsc{LTEprobe}}
\newcommand{\dlsniffer}{\textsc{DOWNLINKprobe}}
\newcommand{\ulsniffer}{\textsc{UPLINKprobe}}

\usepackage{amsmath}
\usepackage{caption}
\usepackage{subcaption}

\usepackage[inline]{enumitem}

\usepackage{fontawesome}

\usepackage{booktabs} 
\newcommand{\ra}[1]{\renewcommand{\arraystretch}{#1}}

\usepackage{makecell}

\makeatletter
\renewcommand\paragraph{\@startsection{paragraph}{4}{\z@}%
                                    {1ex \@plus1ex \@minus.2ex}%
                                    {-1em}%
                                    {\normalfont\normalsize\bfseries}}
\makeatother

\def\enbnotext#1{
\begin{scope}[shift={#1}]
    \draw [very thick]([shift=(-25:0.9)]-.725\unitlen, 1.5\unitlen) arc[start angle=-25, end angle=25, radius=0.9];
    \draw [very thick]([shift=(-25:0.5)]-.725\unitlen, 1.5\unitlen) arc[start angle=-25, end angle=25, radius=0.5];
    \draw [very thick]([shift=(-25:0.7)]-.725\unitlen, 1.5\unitlen) arc[start angle=-25, end angle=25, radius=0.7];
    
    \draw [fill=lgray] (-0.625\unitlen, 0) rectangle (-1.125\unitlen,0.75\unitlen);
    \draw [fill=gray] (-0.795\unitlen,0.75\unitlen) rectangle (-0.955\unitlen,1\unitlen);
    
    \draw [fill=gray] (-0.395\unitlen,1.45\unitlen) rectangle (-1.355\unitlen,1.55\unitlen);
    
    \draw [fill=vlgray] (-0.775\unitlen,1\unitlen) rectangle (-0.975\unitlen,2\unitlen);
    \draw [fill=vlgray] ($(-0.775\unitlen,1\unitlen) + (0.4\unitlen, 0)$) rectangle ($(-0.975\unitlen,2\unitlen) + (0.4\unitlen, 0)$);
    \draw [fill=vlgray] ($(-0.775\unitlen,1\unitlen) - (0.4\unitlen, 0)$) rectangle ($(-0.975\unitlen,2\unitlen) - (0.4\unitlen, 0)$);
    
\end{scope}
}

\def\enbnowaves#1{
\begin{scope}[shift={#1}]
    \node [align=center,text width=2\unitlen, inner sep=0,outer sep=0, minimum height=0.5\unitlen, anchor=south west] at (-0.625\unitlen,0) {\textbf{eNodeB}};

    \draw [fill=lgray] (-0.625\unitlen, 0) rectangle (-1.125\unitlen,0.75\unitlen);
    \draw [fill=gray] (-0.795\unitlen,0.75\unitlen) rectangle (-0.955\unitlen,1\unitlen);
    
    \draw [fill=gray] (-0.395\unitlen,1.45\unitlen) rectangle (-1.355\unitlen,1.55\unitlen);
    
    \draw [fill=vlgray] (-0.775\unitlen,1\unitlen) rectangle (-0.975\unitlen,2\unitlen);
    \draw [fill=vlgray] ($(-0.775\unitlen,1\unitlen) + (0.4\unitlen, 0)$) rectangle ($(-0.975\unitlen,2\unitlen) + (0.4\unitlen, 0)$);
    \draw [fill=vlgray] ($(-0.775\unitlen,1\unitlen) - (0.4\unitlen, 0)$) rectangle ($(-0.975\unitlen,2\unitlen) - (0.4\unitlen, 0)$);
    
\end{scope}
}

\def\enbnowavesnotext#1{
\begin{scope}[shift={#1}]

    \draw [fill=lgray] (-0.625\unitlen, 0) rectangle (-1.125\unitlen,0.75\unitlen);
    \draw [fill=gray] (-0.795\unitlen,0.75\unitlen) rectangle (-0.955\unitlen,1\unitlen);
    
    \draw [fill=gray] (-0.395\unitlen,1.45\unitlen) rectangle (-1.355\unitlen,1.55\unitlen);
    
    \draw [fill=vlgray] (-0.775\unitlen,1\unitlen) rectangle (-0.975\unitlen,2\unitlen);
    \draw [fill=vlgray] ($(-0.775\unitlen,1\unitlen) + (0.4\unitlen, 0)$) rectangle ($(-0.975\unitlen,2\unitlen) + (0.4\unitlen, 0)$);
    \draw [fill=vlgray] ($(-0.775\unitlen,1\unitlen) - (0.4\unitlen, 0)$) rectangle ($(-0.975\unitlen,2\unitlen) - (0.4\unitlen, 0)$);
    
\end{scope}
}

\def\sniffer#1{
\begin{scope}[shift={#1}]
    \draw [fill=lgray] (-1\unitlen, 0) rectangle (1\unitlen,0.5\unitlen);
    \draw [dred, fill=dred] (-0.75\unitlen, 0.25\unitlen) circle (0.05\unitlen);
    \draw [dred, fill=orange] (-0.45\unitlen, 0.25\unitlen) circle (0.05\unitlen);
    
    \draw [fill=black] (0.7\unitlen, 0.225\unitlen) rectangle (0.8\unitlen, 1.5\unitlen);
    \draw [fill=black] (0.715\unitlen, 0.225\unitlen) rectangle (0.785\unitlen, 1.75\unitlen);
    
    \draw [fill=black] (0.4\unitlen, 0.225\unitlen) rectangle (0.5\unitlen, 1.5\unitlen);
    \draw [fill=black] (0.415\unitlen, 0.225\unitlen) rectangle (0.485\unitlen, 1.75\unitlen);

\end{scope}
}

\def\ue#1{
\begin{scope}[shift={#1}]
    \draw [fill=gray] (0, 0) rectangle (1\unitlen,1.3);
    \draw [fill=vlgray] (0.1\unitlen, 0.4) rectangle (0.9\unitlen,1.2);
    \filldraw[fill=vlgray](0.5\unitlen,0.2) circle (0.1);
    
    \node [align=center, inner sep=0,outer sep=0, minimum height=1\unitlen] at (0.5\unitlen,0.8) {\textbf{UE}};
\end{scope}
}

\def\uenotext#1{
\begin{scope}[shift={#1}]
    \draw [fill=gray] (0, 0) rectangle (1\unitlen,1.3);
    \draw [fill=vlgray] (0.1\unitlen, 0.4) rectangle (0.9\unitlen,1.2);
    \filldraw[fill=vlgray](0.5\unitlen,0.2) circle (0.1);
    
\end{scope}
}

\def\uenotextyellow#1{
\begin{scope}[shift={#1}]
    \draw [fill=red, very thick] (0, 0) rectangle (1\unitlen,1.3);
    \draw [fill=vlgray, thick] (0.1\unitlen, 0.4) rectangle (0.9\unitlen,1.2);
    \filldraw[fill=vlgray, thick](0.5\unitlen,0.2) circle (0.1);
    
\end{scope}
}

\def\uenotextopaque#1{
\begin{scope}[shift={#1}]
    \draw [fill=gray, opacity=0.2] (0, 0) rectangle (1\unitlen,1.3);
    \draw [fill=vlgray, opacity=0.2] (0.1\unitlen, 0.4) rectangle (0.9\unitlen,1.2);
    \filldraw[fill=vlgray, opacity=0.2](0.5\unitlen,0.2) circle (0.1);
    
\end{scope}
}

\def\sniffertext#1{
\begin{scope}[shift={#1}]
    \sniffer{(0,0)}
    \node [align=center,text width=4\unitlenbox, anchor=north, inner sep=0,outer sep=0, minimum height=1\unitlen] at (0,0) {\bf \uldlsniffer{}};
\end{scope}
}

\begin{document}
%-------------------------------------------------------------------------------

%don't want date printed
\date{}

\title{\Large \bf \textsc{LTrack}: Stealthy Tracking of Mobile Phones in LTE}

%for single author (just remove % characters)
% \author{\phantom{abc}}
\author{
{\rm Martin Kotuliak, Simon Erni, Patrick Leu, Marc Röschlin, and Srdjan \v{C}apkun}\\
ETH Zurich
}
\maketitle

%-------------------------------------------------------------------------------
\begin{abstract}
%-------------------------------------------------------------------------------

We introduce \textsc{LTrack}, a new tracking attack on LTE that allows an attacker to stealthily extract user devices' locations and permanent identifiers (IMSI). To remain stealthy, the localization of devices in \textsc{LTrack} is fully passive, relying on our new uplink/downlink sniffer. Our sniffer records both the times of arrival of LTE messages and the contents of the Timing Advance Commands, based on which \textsc{LTrack} calculates locations. \textsc{LTrack} is the first to show the feasibility of a passive localization in LTE through implementation on software-defined radio.

Passive localization attacks reveal a user's location traces but can at best link these traces to a device's pseudonymous temporary identifier (TMSI), making tracking in dense areas or over a long time-period challenging. \textsc{LTrack} overcomes this challenge by introducing and implementing a new type of IMSI Catcher named IMSI Extractor. It extracts a device's IMSI and binds it to its current TMSI. Instead of relying on fake base stations like existing IMSI Catchers, which are detectable due to their continuous transmission, IMSI Extractor relies on our uplink/downlink sniffer enhanced with surgical message overshadowing. This makes our IMSI Extractor the stealthiest IMSI Catcher to date.

We evaluate \textsc{LTrack} through a series of experiments and show that in line-of-sight conditions, the attacker can estimate the location of a phone with less than 6m error in 90\% of the cases. We successfully tested our IMSI Extractor against a set of 17 modern smartphones connected to our industry-grade LTE testbed. We further validated our uplink/downlink sniffer and IMSI Extractor in a test facility of an operator.

\end{abstract}

%-------------------------------------------------------------------------------
\section{Introduction}
\label{sec:introduction}
%-------------------------------------------------------------------------------

LTE is one of the most widely deployed and used cellular technologies. It was designed to not only enable communication but also to protect the security and privacy of users by encrypting communication between a user equipment (UE) and a base station (eNodeB). Unlike the user's data, LTE physical and MAC layer control messages are transmitted in plain-text, with subscriber identifiers (IMSI) replaced with temporary identifiers (TMSI) to protect users' privacy.

LTE security and specifically the security and privacy on the wireless link between base stations and UEs is an active area of research. Broadly, attacks against LTE can be classified as active or passive, where active attacks (e.g., IMSI Catcher~\cite{shaik_practical_2017, jover_lte_2016}) typically rely on fake base stations to which victim UEs connect. 
Recently, message overshadowing emerged as a new active, but stealthier manipulation technique~\cite{yang_hiding_2019,erni_adaptover_2021}.

On the other hand, passive attacks rely on custom-built sniffers. In~\cite{kumar_lte_2014, bui_owl_2016}, it was shown that an attacker can build a passive downlink traffic sniffer (from the eNodeB to the UE) using software-defined radios. Downlink sniffers were then used as tools for localization~\cite{roth_location_2017}, to break the encryption of phone calls~\cite{rupprecht_eavesdropping_2020}, and to allow traffic fingerprinting~\cite{kohls_lost_2019}. 
The idea of passive uplink and downlink sniffing was further proposed for user localization~\cite{roth_location_2017} but was not implemented. 
Unlike downlink sniffing, so far, uplink sniffing was implemented only using active techniques and relied on fake base stations~\cite{shaik_new_2019, rupprecht_breaking_2019}. 

In this work, we focus on large-scale, stealthy UE tracking. To be successful in such an attack, the adversary needs to: \begin{enumerate*}[label=(\roman*)]
\item determine the location of the UE, 
\item obtain a UE's identifier that links observed locations into a trace, and 
\item avoid detection. \end{enumerate*}
Until now, no attack fulfills all of the above at the same time.
Passive localization alone could leak UE traces in some low-density areas, but in urban areas with a high density of UEs, this task will be harder without the identifier that binds the observed locations together ~\cite{shokri_quantifying_2011}. 

IMSI Catchers, which are used to leak a UE's IMSI to the adversary and therefore identify the UE, rely exclusively on fake base stations. However, to get the UE to connect to the fake base station (a requirement of the attack), the attacker needs to transmit continuously at a high power and can therefore be detected by law enforcement and operators~\cite{nakarmi_prajwol_kumar_3gpp_2019, quintin_detecting_2020, li_fbs-radar_2017}. 

This paper addresses the above and shows that stealthy localization and identification (and therefore tracking) of UEs in LTE is indeed possible. We present \textsc{LTrack}, a new tracking attack on LTE which combines passive and stealthy active attacks. For passive localization, we use \uldlsniffer{}, our uplink/downlink sniffer, and for binding the collected traces to an IMSI, we use our active but stealthy IMSI Extractor.

Our work focuses on the recovery of users' long-term mobility traces. How this information is then further used by adversaries is well studied and out of scope of our work. Prior research showed that traces can be used to deanonymize users through transportation routines~\cite{liao_learning_2007}, mobility patterns~\cite{wang_-anonymization_2018, pyrgelis_knock_2018, zang_anonymization_2011, gambs_-anonymization_2014}, home addresses~\cite{hoh_enhancing_2006, lamarca_inference_2007, tokuda_anonymity_2009}, co-locations with other users~\cite{olteanu_quantifying_2014, srivatsa_deanonymizing_2012}, or online geo-tagged media~\cite{henne_snapme_2013}.

In summary, we make the following contributions:

\begin{itemize}
   \item We demonstrate the feasibility of a fully passive adversarial localization of UEs in an LTE network. We show that, in line-of-sight conditions, the attacker can estimate the location of a phone with an error of less than 6m in 90\% of the cases.
   
   \item We propose a new type of IMSI Catcher, named IMSI Extractor. Our IMSI Extractor does not rely on fake base stations but instead uses a combination of low-power surgical message overshadowing and uplink/downlink sniffing. Even if our catcher injects a message, it does so in line with LTE protocol specification, making it hard to detect with existing IMSI Catcher detection techniques. We discuss the techniques that would be needed to detect this attack. We successfully tested our IMSI Extractor on 17 smartphones connecting to an industry-grade eNodeB. 
   
   \item We combine our passive localization and our IMSI Extractor into a UE tracking system that we name \textsc{LTrack}, which enables simultaneous identification and localization of UEs, allowing an attacker to track users more persistently and with higher accuracy than in prior attacks. \textsc{LTrack} does this by cross-checking IMSI-TMSI pairs obtained with our IMSI Extractor, with the location data identified by the TMSI obtained from our localization attacks. 
   
   \item We implement the first white-box uplink and downlink LTE sniffer, called \uldlsniffer{}. So far, only downlink sniffers were presented in open research. This sniffer is one of the core components of \textsc{LTrack}. Our sniffer records both protocol level information, e.g., synchronization parameters or phone model specific messages, and physical layer timings of messages. 
   
   \item Using our sniffer, we implement mobile phone fingerprinting, which allows the attacker to identify the make and the model of the phone. This allows us, in some scenarios, to further increase the accuracy of phone localization and tracking by as much as 20 meters. 
   \end{itemize} 

%-------------------------------------------------------------------------------
\section{Background}
\label{sec:background}
%-------------------------------------------------------------------------------

\subsection{LTE}

The radio access network in LTE is managed by base stations (eNodeB). eNodeBs route the traffic over a secure channel to the network core, which handles most mobile network functions. Our sniffer captures and analyzes the communication between a base station and a mobile phone (UE): downlink from eNodeB to UE, and uplink from UE to eNodeB. Most providers implement uplink and downlink separation using FDD-LTE (Frequency Division Duplex). In FDD, uplink and downlink use two separate RF carriers, one for each direction. Multiplexing is implemented using OFDMA in downlink and SC-FDMA in uplink.

Physical layer data transmission is scheduled in 10ms long frames for both downlink and uplink~\cite{3gpp.36.211}. Frames are indexed from 0 to 1023 and split into ten subframes, each with a duration of 1ms. Each subframe consists of two slots. By default, a slot is made up of 7 OFDM symbols with one cyclic prefix per symbol.

In both OFDMA and SC-FDMA, data is modulated onto orthogonal subcarriers. Modulated data values are called frequency samples. Using inverse fast Fourier transformation, frequency samples are transformed into a time signal and transmitted over the radio. An LTE receiver samples the incoming signal into time domain samples. Fast Fourier transform over the time samples outputs the frequency samples. The smallest indexed element is a resource block~\cite{3gpp.36.211} which spans 12 subcarriers and lasts one slot. 

\paragraph{Physical Layer Channels.}

Data on the physical layer is sent over different channels~\cite{3gpp.36.211}. Each channel occupies predefined resource blocks. Physical shared channels are used for data transmission, and control channels manage flow and access to them. The Physical Random Access Channel is used to establish new UE connections.

All resource allocations of resource blocks are communicated to the UE in Downlink Control Information (DCI) elements transmitted over the downlink control channel. A 16-bit RNTI number addresses each DCI and specifies the recipient of the message. Depending on the function, the RNTI number specifies one UE or multiple UEs. The format of the DCI determines its function.

\begin{description}[itemsep=0.1cm, parsep=0pt]
\item[DCI Format 0] allocates resource blocks on the uplink to UEs. A UE can transmit on the uplink shared channel only if it receives a corresponding resource allocation. The DCI Format 0 also specifies parameters to be used for the message encoding, such as modulation schemes.

\item[DCI Format 1 or 2] defines which resource blocks a UE should decode and which parameters it should use to decode the messages on the downlink shared channel. The downlink shared channel carries user data and other system information, such as the configuration of the base station.
\end{description}

\paragraph{Connection Establishment.}

A UE uses two numbers for identifying to the network: IMSI, a unique, persistent identifier, and TMSI, a temporary identifier. Each UE connection starts with an RRC Connection Request containing the UE TMSI. If the TMSI is not available, the UE samples a random value and includes it instead of the TMSI.

There are two ways how a UE requests the service from the network. If the UE connects for the first time after losing a state (e.g., restarting), it initiates an attachment procedure by sending an Attach Request, containing the TMSI if one has been assigned previously, or the IMSI otherwise. If the network does not recognize the TMSI, it will ask the UE to provide its IMSI in an identification procedure. At the end of the attachment procedure, after the security context has been set up, the network assigns the UE a new TMSI. The TMSI is at this point both ciphered and integrity protected.

If the UE is already attached to the network but idle, going from idle state to connected state, it enters the service request procedure by sending an integrity protected Service Request, after which the connectivity is immediately restored. 

\subsection{Relevant Attacks}

\paragraph{Localization Attacks.}
By observing paging messages alone, an attacker can learn if a victim is currently in the same tracking area or the same cell (if smart paging is deployed), as shown in \cite{shaik_practical_2017}.

With the victim connected to the same base station as the attacker, more advanced attacks can be executed. As proposed in \cite{roth_location_2017}, an attacker can observe control messages on the MAC layer that contain propagation delay correction information. This information alone constrains the location of the victim to a 78 meters wide ring around the eNodeB with its perimeter defined by the propagation delay correction.

Localization attacks based on fake base stations~\cite{jover_lte_2016} are even more accurate. However, we do not consider them stealthy enough to be used in a large-scale tracking attack. 

\paragraph{IMSI Catchers.}

As mentioned in \autoref{sec:introduction}, for areas with a high density of UEs, the attacker needs to be able to obtain the identity of the victims in order to track them. The most potent attacks in this area are IMSI Catchers~\cite{shaik_practical_2017, jover_lte_2016}, which reveal the unique IMSI number to the attacker. However, these attacks all rely on fake base stations.

%-------------------------------------------------------------------------------
\section{\uldlsniffer}
\label{sec:sniffer}
%-------------------------------------------------------------------------------

The key component to make the stealthy tracking possible is the implementation of a combined uplink/downlink LTE sniffer that we name \uldlsniffer{}. In what follows, we describe \uldlsniffer{} and its abilities. As already discussed, downlink sniffing (see, e.g., \cite{kumar_lte_2014, bui_owl_2016}) allows the attacker to record unencrypted Downlink Control Information and control elements on MAC layer. 
With an uplink sniffer, however, the attack surface increases substantially. Unencrypted messages, such as the initialization messages, can be used by the attacker for the leakage of users' identifiers. All uplink messages, even encrypted ones, can be used for precise time of arrival measurements.

\begin{figure*}
    \begin{center}
    \begin{tikzpicture}[remember picture, 
    font=\small\ttfamily]
    
\definecolor{vlgray}{rgb}{0.9, 0.9, 0.9}
\definecolor{lgray}{rgb}{0.8, 0.8, 0.8}
\definecolor{gray}{rgb}{0.6, 0.6, 0.6}
\definecolor{dgray}{rgb}{0.5, 0.5, 0.5}
\definecolor{vdgray}{rgb}{0.2, 0.2, 0.2}

\setlength{\unitlen}{0.7cm}
\setlength{\unitlenbox}{0.7cm}

%\clip (-14.5,-2.75) rectangle (3.6,5.42);

%Place a `comment` node with all text inside a tabular  
\node(comments) {\normalfont \begin{tabular}{l@{ }l}
        \subnode{A0}& \\[2.5mm]
        \subnode{B0}{\uldlsniffer{} parses \texttt{RAR} and gets an \texttt{RNTI} of a newly }& \\ 
        {connected UE; \uldlsniffer{} can decode \texttt{DCI}s using the \texttt{RNTI}} &\\[5mm] 
        \subnode{A}{\uldlsniffer{} uses a \texttt{DCI Format 1} to decode downlink \texttt{DATA$_\texttt{1}$}}& \\[1mm]
        \subnode{B}{}& \\[2.5mm]
        \subnode{C}{\uldlsniffer{} combines \texttt{DATA$_\texttt{1}$}, \texttt{DATA$_\texttt{2}$}, ... \texttt{DATA$_\texttt{n}$} to get }& \\ 
        a higher layer packet &\\[5mm] 
        \subnode{D}{\uldlsniffer{} and UE receive a \texttt{DCI Format 0} }&  \\ 
        with an uplink scheduling information&\\
        \subnode{E}{\uldlsniffer{} uses the scheduling from the \texttt{DCI Format 0}}& \\
        to decode uplink \texttt{DATA} and measure the time of arrival &\\
        \end{tabular}};

% Place `server` and `client` nodes
\node[above left=5mm and 0mm of comments, anchor=center] (ue) {\normalfont \textbf{Mobile Phone}};
\node[left=15mm of ue, red, anchor=center] (sniffer) {\normalfont \textbf{\uldlsniffer{}}}; 
\node[left=50mm of sniffer, anchor=center] (enb) {\normalfont \textbf{Base Station}}; 
% \node[above=-1mm of enb] {\normalfont \textbf{Base Station}}; 
\enbnowavesnotext{($(enb) + (0.6cm, 0.3cm)$)};
\sniffer{($(sniffer) + (0, 0.3cm)$)}
\uenotext{($(ue) + (-0.35cm, 0.3cm)$)}

\definecolor{vlgray}{rgb}{0.9, 0.9, 0.9}
\definecolor{lgray}{rgb}{0.8, 0.8, 0.8}

% \filldraw [fill=vlgray, draw=vlgray] ($ (D2-|ue2) + (-4mm, 2mm) $) rectangle ($ (F2-|epc2) + (2mm, 6mm) $);
% \filldraw [fill=vlgray, draw=vlgray] ($ (H2-|ue2) + (-4mm, 4mm) $) rectangle ($ (comments2.south-|epc2) + (2mm, 0) $);
% \filldraw [fill=lgray, draw=lgray] ($ (H2-|ue2) + (-2mm, 2mm) $) rectangle ($ (comments2.south-|enb2) + (2mm, 0) $);

%Draw vertical lines
\draw [thick]($(ue)+(0,-0.5cm)$) -- ($(ue|-comments.south)+(0,0.5cm)$);
\draw [thick]($(enb)+(0,-0.5cm)$) -- ($(enb|-comments.south)+(0,0.5cm)$);
\draw [red,thick, dotted] ($(sniffer)+(0,-0.5cm)$) -- ($(sniffer|-comments.south)+(0,0.5cm)$);

%Draw message interchanges
\draw[->,thick] ($(A0-|ue) + (-1mm, 0)$)-- (A0-|sniffer) -- node[below]{Random Access Preamble} ($(A0-|enb) + (1mm, 0)$) ;
\draw[<-,thick] ($(B0-|ue) + (-1mm, 0)$)-- (B0-|sniffer) -- node[below]{Random Access Response} ($(B0-|enb) + (1mm, 0)$);
\draw[<-,thick] ($(A-|ue) + (-1mm, 0)$)-- (A-|sniffer) -- node[below]{DCI Format 1, DATA$_\texttt{1}$} ($(A-|enb) + (1mm, 0)$);
\draw[<-,thick] ($(B-|ue) + (-1mm, 0)$)-- (B-|sniffer) -- node[below=1.5mm]{...} ($(B-|enb) + (1mm, 0)$);
\draw[<-,thick] ($(C-|ue) + (-1mm, 0)$)-- (C-|sniffer) -- node[below]{DCI Format 1, DATA$_\texttt{n}$} ($(C-|enb) + (1mm, 0)$);
\draw[<-,thick] ($(D-|ue) + (-1mm, 0)$)-- (D-|sniffer) -- node[below]{DCI Format 0} ($(D-|enb) + (1mm, 0)$) ;
\draw[->,thick] ($(E-|ue) + (-1mm, 0)$)-- (E-|sniffer) -- node[below]{Uplink DATA} ($(E-|enb) + (1mm, 0)$) ;

\filldraw [fill=white, draw=white] ($ (A-|enb) + (-4mm, 2mm) $) rectangle ($ (A-|ue) + (2mm, 6mm) $);
\filldraw [fill=white, draw=white] ($ (D-|enb) + (-4mm, 2mm) $) rectangle ($ (D-|ue) + (2mm, 6mm) $);

\node[red, fill=white] at ($(B0 -| sniffer)$){\faEye};
\node[red, fill=white] at ($(A -| sniffer)$){\faEye};
\node[red, fill=white] at ($(B -| sniffer)$){\faEye};
\node[red, fill=white] at ($(C -| sniffer)$){\faEye};
\node[red, fill=white] at ($(D -| sniffer)$){\faEye};
\node[red, fill=white] at ($(E -| sniffer)$){\faEye};

\node [anchor=south, rotate=90, text width=1.5cm, align=center] at ($(E -| enb)+(0,0.4cm)$) {\normalfont Uplink Sniffing};

\node [anchor=south, rotate=90, text width=2cm, align=center] at ($(B -| enb)+(0,-0.2cm)$) {\normalfont Downlink Sniffing};

\node [anchor=south, rotate=90, text width=1cm, align=center] at ($(B0 -| enb)+(0,0.2cm)$) {\normalfont RNTI Parsing};

\end{tikzpicture}
    \end{center}
    \caption{Decoding of uplink and downlink channels by \uldlsniffer{}. First \uldlsniffer{} records an RNTI of a UE. Then it uses it to decode DCIs. DCIs either specify resource blocks containing downlink or uplink data.}
    \label{fig:ul_dl_sniffer}
\end{figure*}
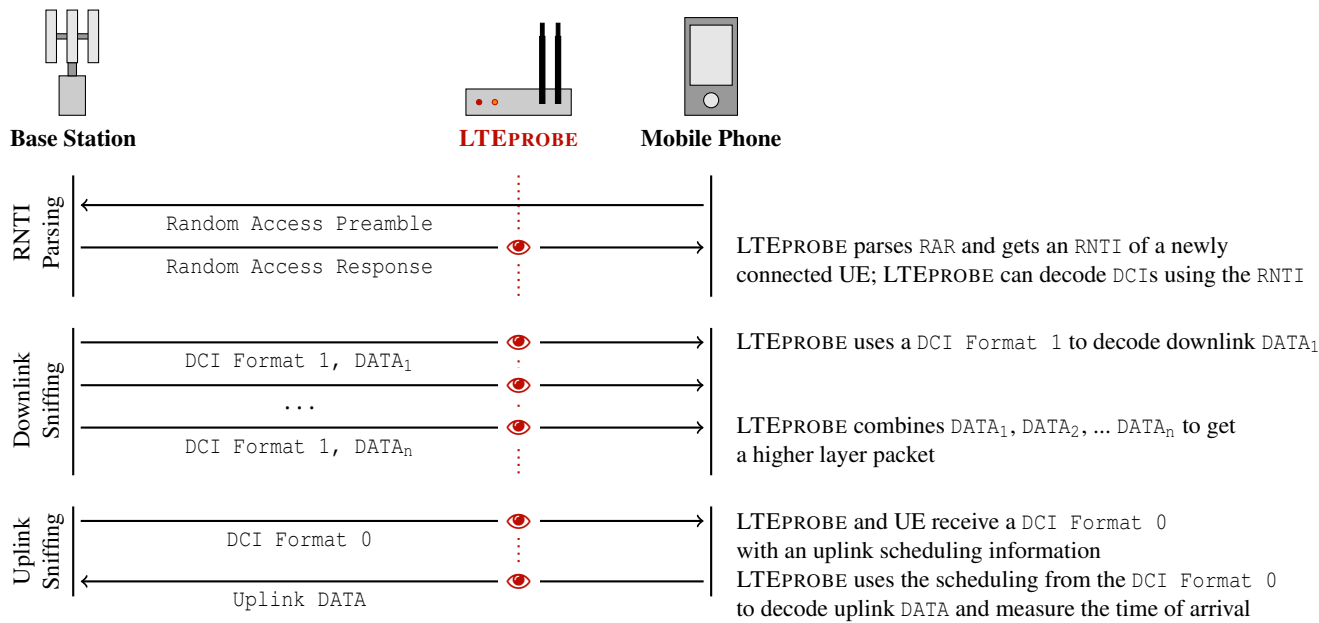

\subsection{System Architecture}

We designed \uldlsniffer{} to be a fully passive device and therefore virtually undetectable. \uldlsniffer{} receives RF samples on both uplink and downlink. It records all communication between mobile phones and base stations but does not break encryption. \uldlsniffer{} has a stable clock and synchronizes its reception to the base station. 
The clock drift between the base station's and \uldlsniffer's clock is negligible, because both devices use GPS synchronized clocks.

\uldlsniffer{} consists of two components: \dlsniffer{}, the downlink sniffer, and \ulsniffer{}, the uplink sniffer. \dlsniffer{} works as a standalone analyzer for downlink, but \ulsniffer{} requires scheduling information shared by \dlsniffer{}. The uplink and downlink sniffing is feasible due to the unencrypted DCI messages carried over the downlink control channel.

\paragraph{\dlsniffer}
first synchronizes to the base station and records identifiers of connected UEs. The LTE protocol specifies temporary RNTI numbers for the identification of UEs on the physical layer for the duration of the connection. With the RNTI, \dlsniffer{} finds and decodes messages intended for the victim UEs. 

On the physical layer, when a UE connects to an eNodeB, the eNodeB replies with a Random Access Response. Inside the Random Access Response, the eNodeB specifies a new RNTI for the UE. Because the Random Access Response is sent in a plain-text, it is visible to \dlsniffer{}. However, this method can be used only for new connections. For already connected UEs, the assigned RNTI is not exchanged in plain-text, but coded into the CRC of DCI messages. The work in~\cite{kumar_lte_2014} describes a method for extracting the RNTIs from the DCIs, however, for our use-cases this is not necessary.

To decode downlink channels, \dlsniffer{} computes inverse OFDMA transformation to receive frequency samples. It performs channel correction and frequency offset correction. It then goes over all possible locations of DCIs for the set of recorded RNTIs and tries to decode them. Depending on the format of the decoded DCI, \dlsniffer{} either uses it to decode the PDSCH message or shares it with \ulsniffer{}.
Finally, \dlsniffer{} parses PDSCH messages to get higher-layer messages, e.g., NAS-layer messages containing dedicated UE configuration. \autoref{fig:ul_dl_sniffer} shows how \dlsniffer{} receives RNTIs in Random Access Response and then obtains shared channel data.

\paragraph{\ulsniffer}
receives samples transmitted from multiple UEs. Similar to the eNodeB, it demodulates them and applies channel correction. Afterward, it tries to decode uplink shared channels and control channels.

The physical uplink shared channel is decoded according to the scheduling information. eNodeB controls the scheduling in the LTE protocol, so it knows the scheduled resource allocations. In our case, \ulsniffer{} has to obtain the scheduling information from the DCI messages the same way the UE receives them. \ulsniffer{} uses the passed DCI Format 0 messages from the downlink sniffer containing the scheduling information. Because DCI Format 0 messages carry the resource allocations for future uplink transmissions, without a downlink sniffer, \ulsniffer{} would not be able to decode uplink channels. \autoref{fig:ul_dl_sniffer} visualizes the procedure of \ulsniffer{}. 

To correctly decode uplink shared and control channels, \ulsniffer{} has to apply a dedicated UE configuration sent via a RRC-layer downlink message. \ulsniffer{} again uses information recorded by \dlsniffer{}. Similar to \dlsniffer{}, physical layer messages are parsed to receive higher layer messages.

\subsubsection*{\uldlsniffer{} Implementation}\label{sec:sync}

We base our implementation on srsLTE~\cite{gomez-miguelez_srslte_2016}, an open-source library for the LTE protocol. The two main components, \dlsniffer{} and \ulsniffer{}, run on two separate co-located USRP devices. The two components run as two threads of a parent \uldlsniffer{} program.

Both downlink and uplink subframes are scheduled at the same time. Regular UEs learn the timings of the subframe from the synchronization signals transmitted by the eNodeB. Similarly, our \uldlsniffer{} synchronizes to the eNodeB by observing the synchronization signals. However, only \dlsniffer{} is receiving them. Therefore, the uplink and downlink threads of \uldlsniffer{} need to share timings and subframe numbers; otherwise, \ulsniffer{} would not be able to receive the uplink subframes at the correct time. To synchronize precisely, the two USRPs need to have the same time reference. This can be solved by using a GPSDO on both USRPs to have the same GPS clock or by using an Octoclock, a clock distribution module. 

For each subframe, both \ulsniffer{} and \dlsniffer{} record the subframe index and the exact time they received it. If the timestamps for the same subframe index do not match, \ulsniffer{} has to adjust its reception time by discarding time samples. A perfect synchronization of the two components is then achieved.

Unless \uldlsniffer{} is co-located with the eNodeB, the uplink messages sent by UEs will not be perfectly time synchronized to the frames at \uldlsniffer{} location (due to different propagation delays). We tested the robustness of the \uldlsniffer{} under such misalignment of the attacker in \autoref{fig:delay} in the Appendix. Our results show that the attacker can still decode the messages under $< 4\mu s$ misalignment (which corresponds to 1.2$km$ distance). However, even if this misalignment would be larger, because \ulsniffer{} and \dlsniffer{} are independent devices, the attacker can apply time correction to the uplink messages and still correctly decode uplink messages.

%-------------------------------------------------------------------------------
\section{Passive Localization Attack}
\label{sec:localization}
%-------------------------------------------------------------------------------

\begin{figure}
    \begin{center}
    \begin{tikzpicture}[remember picture, 
    font=\small\ttfamily]
    
      \definecolor{vlgray}{rgb}{0.9, 0.9, 0.9}
      \definecolor{lgray}{rgb}{0.8, 0.8, 0.8}
      \definecolor{gray}{rgb}{0.6, 0.6, 0.6}
      \definecolor{dgray}{rgb}{0.5, 0.5, 0.5}
      \definecolor{vdgray}{rgb}{0.2, 0.2, 0.2}

\tikzset{
  unit length/.code={\setlength{\unitlen}{#1}},
  unit length = 0.7cm,
  node distance = 2,
  auto,
  on grid,
}

\tikzset{
  unit length/.code={\setlength{\unitlenbox}{#1}},
  unit length = 0.7cm,
  node distance = 2,
  auto,
  on grid,
}

\clip (-5,-1.7) rectangle (3.6,7.1);

% \draw[fill=yellow,even odd rule]  (0,0) circle (3)
%                                   (0,0) circle (2);
                                   
% \draw[fill=blue,even odd rule, fill opacity = 0.5]  (2,0) circle (3)
%                                   (2,0) circle (2);
% \enb{(5,0\unitlen)}

% \ue{(2,2\unitlen)}

\def\ringa{(0,0) circle (3) (0,0) circle (5)}
\def\ringb{(3,0) ellipse (4.5 and 3.35) (3,0) ellipse (5 and 4)}
\def\ringc{(-2,2.5) ellipse (4.5 and 3.35) (-2,2.5) ellipse (5 and 4)}

\fill [fill=lyellow, even odd rule] \ringa;
\fill [fill=vlgray, even odd rule, rotate around={-45:(-2,2.5)}] \ringc;

\begin{scope}[even odd rule]
    % Define a clipping path. All paths outside ringa will
    % be cut because the even odd rule is set. 
    \clip \ringa;
    % Fill ringb. Since the even odd rule is set, only the
    % ring will be filled, not the hole in the middle.  
    \fill[fill=yellow, rotate around={-45:(-2,2.5)}] \ringc;
\end{scope}

% Then we draw the rings

\uenotextopaque{(1.2, 3.2)}
\uenotextopaque{(0.6, 3.9)}

\uenotextopaque{(-5, -0.5)}
\uenotextopaque{(-4.4, -1)}
\uenotextopaque{(-3.6, -1.7)}

\ue{(1.8, 2.7)}

\enbnowaves{(0.5,-0.5)}
%\sniffer{(6,0)}
\sniffertext{(-3.8,4.5)}

\draw[<->, thick] (0.4,0.3) -- node[below, anchor=north, rotate=65]{\normalfont{TA Command}} (1.7,3) ;

\draw[<->, thick] (0.4,0.8) -- (1.7,3.5) -- node[below, text width=4cm, align=center, anchor=north, rotate=-20]{$\delta_{\operatorname{UE}} + \delta_{\operatorname{\normalfont{\textsc{ULprobe}}}}$} (-3, 5.3) ;
    
\end{tikzpicture}
    \end{center}
    \caption{Passive localization attack using a single sniffer. The yellow ring is defined by the received Timing Advance Command, and the grey ellipse is defined by the time of arrival measured by \uldlsniffer{}. The intersection of the two rings defines possible locations of the mobile phone.}
    \label{fig:localization_attack_single}
\end{figure}
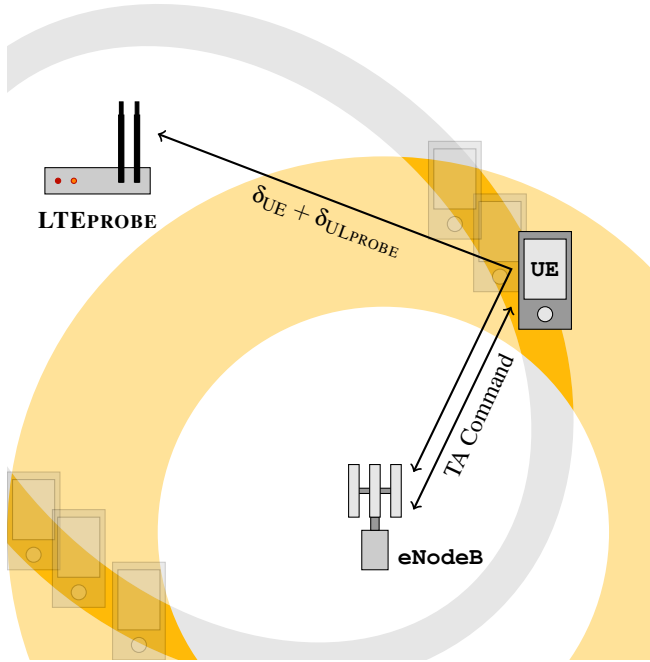

Passive localization of UEs in LTE networks was proposed in a number of prior works~\cite{kumar_lte_2014,shaik_practical_2017,roth_location_2017}. Most notably, Roth et al.~\cite{roth_location_2017} proposed a passive localization attack that leverages synchronization parameters sent on the MAC layer (Timing Advance Command) and times of arrival of uplink and downlink messages.

Specifically, the attack proposed in \cite{roth_location_2017} works by observing the Timing Advance Command containing propagation delay correction information. Because of the coarse granularity of the Timing Advance Command, the attack constrains the location of the victim to a 78 meters wide ring around the eNodeB. Furthermore, in LTE-Advanced, the UE has an option to connect to multiple cells at once. Multiple delay correction information then constrains the victim's location to the intersection of the rings. Finally, Roth et al.~\cite{roth_location_2017} proposes an idea of localizing a UE based on times of arrival of uplink messages, which would allow the attacker to constrain the victim's location to an additional 78 meters wide ring around the attacking device. However, the authors do not provide details, simulations, or implementation of this proposal.The LTE Positioning Protocol~\cite{3gpp.36.355} has been standardized/implemented for localization in LTE. One supported technology is estimating the location from the UE with the observed time difference of arrival (OTDOA) of downlink transmission from multiple base station in the vicinity. The mechanism of the OTDOA method is the same as the one proposed by Roth et al.

Our passive localization attack also exploits unciphered Timing Advance Command and time of arrival of uplink messages. However, contrary to their work, we transform the geometry of the problem from a circle to an ellipse. This transformation enables us to remove the systematic error due to the course-grained Timing Advance Command. Instead of having an additional 78m wide ring around the sniffer, we have a precise ellipse with focal points at the base station and the sniffer, as drawn in \autoref{fig:localization_attack_single}.

The most significant contribution of our attack is the actual implementation and its evaluation in \autoref{sec:results}. Our implementation revealed the imprecision of the hardware inside the mobile phones. To solve this problem, we have transformed the active fingerprinting attack introduced in \cite{shaik_new_2019} into an entirely passive attack in \autoref{sec:fingerprinting}. Knowing the model of mobile phones can increase the precision of the localization by as much as 20 meters. In our work, we do not study the effects of the radio environment (e.g., multi-path propagation or shadow-fading) as these topics are orthogonal to our research. For performance reasons, any positioning system has to account for such conditions. The work in~\cite{sven_fischer_observed_nodate} provides an example how one can use the error budget to compute the precision of localization system under different channel conditions.

In this section, we develop this passive localization attack, provide its mathematical basis and describe its implementation. The attacker can constrain the victim's location to two possible areas as shown in \autoref{fig:localization_attack_single} using just one sniffing device and a base station. The two possible areas are the intersection of a wide ring defined by the Timing Advance Command and an ellipse defined by the time of arrival of uplink messages. Using two or more sniffing devices results in the attacker learning the location of the victim. Alternatively, the adversary can rule out possible locations by cross-checking with, e.g, a detailed map of the area.

\subsection{Timing Advance Command}\label{sec:synchronization}

\begin{figure}
    \centering
    \begin{tikzpicture}[remember picture, 
    font=\small]
    
      \definecolor{vlgray}{rgb}{0.9, 0.9, 0.9}
      \definecolor{lgray}{rgb}{0.8, 0.8, 0.8}
      \definecolor{gray}{rgb}{0.6, 0.6, 0.6}
      \definecolor{dgray}{rgb}{0.5, 0.5, 0.5}
      \definecolor{vdgray}{rgb}{0.2, 0.2, 0.2}

\tikzset{
  unit length/.code={\setlength{\unitlen}{#1}},
  unit length = 0.7cm,
  node distance = 2,
  auto,
  on grid,
}

\tikzset{
  unit length/.code={\setlength{\unitlenbox}{#1}},
  unit length = 0.7cm,
  node distance = 2,
  auto,
  on grid,
}

\clip (-0.1, 3.7) rectangle (8.2,-2.3);

\fill[fill=vlgray] (5.4, 2.2) rectangle (6.9, -1.2);
%\node [anchor = north west] at (5.4, 2) {TA = $2\delta$};
\draw [<->,thick] (5.4, 1.7) -- node [above] {TA = $2\delta$} (6.9, 1.7);

\fill[fill=vlgray] (2.8, 2.2) rectangle (5.2, -1.2);
%\node [anchor = north west] at (2.8, 2) {TA = 0};
\draw [<->,thick] (2.8, 1.7) -- node [above] {TA = 0} (5.2, 1.7);

\draw (-0.2,1) -- (7,1);
\node [anchor=west] at (7,1) {eNodeB};

\draw (-0.2,0) -- (7,0);
\node [anchor=west] at (7,0) {UE};

\draw[->] (-0.2,-1) -- (7,-1);
\node [anchor=west] at (7,-1) {$time$};

\draw[dashed] (0,1.5) -- (0, -1.2);
\draw[dashed] (0.5,0.5) -- (0.5, -1.5);
\draw[dashed] (3,1.5) -- (3, -1.2);
\draw[dashed] (4,1.5) -- (4, -1.5);
\draw[dashed] (6,1.5) -- (6, -1.2);

\node[fill=gray, anchor=south west] at (0, 1) {SF 1};
\node[fill=gray, anchor=south west] at (0.5, 0) {SF 1};

\node[fill=yellow, anchor=south west] at (3.5, 0) {SF 2};
\node[fill=yellow, anchor=south west] at (4, 1) {SF 2};

\node[fill=yellow, anchor=south west] at (5.5, 0) {SF 3};
\node[fill=yellow, anchor=south west] at (6, 1) {SF 3};

\node[anchor=north west] at (0.3, -1.5) {Propagation delay};
\node[anchor=north west, text width = 3.5cm] at (3.8, -1.5) {Without TA, uplink messages would arrive delayed};

\draw[<->] (0,0.2) -- node[above]{$\delta$} (0.5, 0.2);

\draw[<->] (3,0.2) -- node[above]{$\delta$} (3.5, 0.2);

\draw[<->] (3,1.2) -- node[above]{$2 \delta$} (4, 1.2);

\node[anchor=north] at (0, -1.1) {$t_1$};
\node[anchor=north] at (3, -1.1) {$t_2$};
\node[anchor=north] at (6, -1.1) {$t_3$};

\node[fill=gray, anchor=south west] at (0, 2.7) {SF};
\node[anchor=south west] at (0.6, 2.7) {Downlink Subframe};
\node[fill=yellow, anchor=south west] at (0, 2.2) {SF};
\node[anchor=south west] at (0.6, 2.12) {Uplink Subframe};
\node[anchor=south west] at (0, 3.2) {$t_n$};
\node[anchor=south west] at (0.6, 3.2) {Tx/Rx Time of Subframe $n$ at eNodeB};
    
\end{tikzpicture}
    \caption{Timing Advance is used to align uplink transmissions. Transmission and reception at eNodeB are synchronized.}
    \label{fig:timing_advance}
\end{figure}
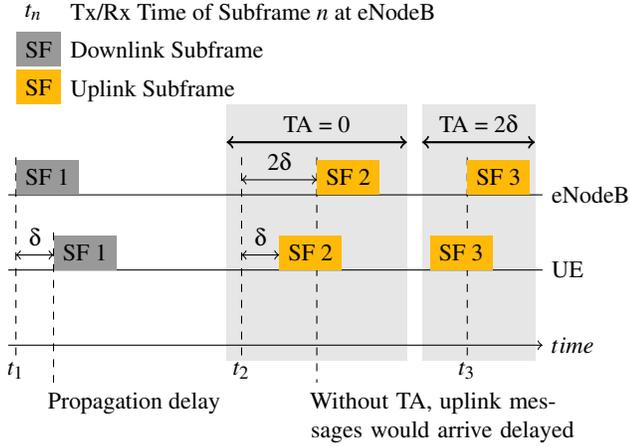

Multiple UEs connect to eNodeB at the same time. Each UE is at a different distance. Due to a propagation delay, without any corrective mechanism, uplink messages would be received with a different delay. Thus, the eNodeB needs to help correct each UE's timing to ensure alignment of all uplink messages within the resource blocks as observed by the eNodeB.

\autoref{fig:timing_advance} shows a situation where the propagation delay between the UE and eNodeB is $\delta$. Due to the propagation delay of the downlink message, the frame synchronization of the UE is being shifted by $\delta$ from the eNodeB's time. The propagation delay of an uplink message is again $\delta$. Therefore, the uplink message arrives at the eNodeB with a delay of $2\delta$. The eNodeB measures the delay and signals it to the UE with a Timing Advance (TA) Command. 

\begin{figure}
    \centering
    \begin{tikzpicture}[remember picture, 
    font=\small]
    
      \definecolor{vlgray}{rgb}{0.9, 0.9, 0.9}
      \definecolor{lgray}{rgb}{0.8, 0.8, 0.8}
      \definecolor{gray}{rgb}{0.6, 0.6, 0.6}
      \definecolor{dgray}{rgb}{0.5, 0.5, 0.5}
      \definecolor{vdgray}{rgb}{0.2, 0.2, 0.2}

\tikzset{
  unit length/.code={\setlength{\unitlen}{#1}},
  unit length = 0.7cm,
  node distance = 2,
  auto,
  on grid,
}

\tikzset{
  unit length/.code={\setlength{\unitlenbox}{#1}},
  unit length = 0.7cm,
  node distance = 2,
  auto,
  on grid,
}

\clip (-0.5,-2) rectangle (6,2);

% \draw[fill=yellow,even odd rule]  (0,0) circle (3)
%                                   (0,0) circle (2);
                                   
% \draw[fill=blue,even odd rule, fill opacity = 0.5]  (2,0) circle (3)
%                                   (2,0) circle (2);
% \enb{(5,0\unitlen)}

% \ue{(2,2\unitlen)}

\def\ringa{(0,0) circle (3) (0,0) circle (5.5)}

\fill [fill=lyellow, even odd rule] \ringa;

\scalebox{0.5}{\uenotext{(6.1, -0.1)}}

\enbnowavesnotext{(0.5,-1)}

\draw[<->, thick] (0.4,-0.5) -- node[below, anchor=north]{1/2 TA Value} (4.25,-0.5);

\draw[<->, thick] (0.4,-0.3) -- node[above]{$\delta$} (3.2,-0.3);

\draw[<->, thick] (3.3,-0.3) -- node[above]{$\epsilon$} (4.25,-0.3);

\draw[<->, thick] (2.5,-1.75) -- node[above]{78m} (5.2,-1.75);
    
\end{tikzpicture}
    \caption{The propagation delay between the eNodeB and UE is $\delta$, but the received Timing Advance value corresponds to a distance in the middle of yellow ring. The difference between these two values is a systematic error $\epsilon$.}
    \label{fig:ta_err}
\end{figure}
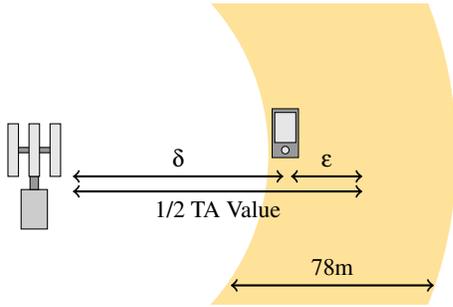

The LTE specification \cite{3gpp.36.214} defines that the Timing Advance value is expressed as $T_A \times 16 \times T_S$, where $T_S = 1/30720ms$. $T_A$ is the value signalled by the eNodeB. The $T_A$ value is sent as a part of the MAC control element. It is sent unciphered by the eNodeB on the MAC layer.

The granularity of the TA is therefore $T_S \times 16 = 0.5208 \mu s$. The UE does not receive a more precise value for the propagation delay $\delta$. Given that the propagation speed is the speed of light, the UE can estimate its distance from the eNodeB in a range of $78.07m$ ($156.14m$ divided by 2 because of the round-trip). \autoref{fig:ta_err} visualizes the difference between the actual distance of the UE and the eNodeB and the distance the UE computes from the TA Command.

\subsection{Times of Arrival of Uplink and Downlink Messages}\label{sec:timings}

\begin{figure*}
    \centering
    \begin{tikzpicture}[remember picture, 
    font=\small]
    
      \definecolor{vlgray}{rgb}{0.9, 0.9, 0.9}
      \definecolor{lgray}{rgb}{0.8, 0.8, 0.8}
      \definecolor{gray}{rgb}{0.6, 0.6, 0.6}
      \definecolor{dgray}{rgb}{0.5, 0.5, 0.5}
      \definecolor{vdgray}{rgb}{0.2, 0.2, 0.2}

\tikzset{
  unit length/.code={\setlength{\unitlen}{#1}},
  unit length = 0.7cm,
  node distance = 2,
  auto,
  on grid,
}

\tikzset{
  unit length/.code={\setlength{\unitlenbox}{#1}},
  unit length = 0.7cm,
  node distance = 2,
  auto,
  on grid,
}

%\clip (-1, -1) rectangle (18,4);

\fill[fill=vlgray] (12, 3.2) rectangle (16, -1.2);
% \node [anchor = north west] at (12, 3) {TA = $2\delta_{\operatorname{UE}} + 2\epsilon$};
\draw [<->,thick] (12, 2.7) -- node [above] {TA = $2\delta_{\operatorname{UE}} + 2\epsilon$} (16, 2.7);

\fill[fill=vlgray] (6.7, 3.2) rectangle (11.8, -1.2);
%\node [anchor = north west] at (6.7, 3) {TA = 0};
\draw [<->,thick] (6.7, 2.7) -- node [above] {TA = 0} (11.8, 2.7);

\draw (-0.2,2) -- (16,2);
\node [anchor=west] at (16,2) {eNodeB};

\draw (-0.2,1) -- (16,1);
\node [anchor=west] at (16,1) {UE};

\draw (-0.2,0) -- (16,0);
\node [anchor=west] at (16,0) {\uldlsniffer{}};

\draw[->] (-0.2,-1) -- (16,-1);
\node [anchor=west] at (16,-1) {$time$};

\draw[dashed] (0,2.5) -- (0, -1.2);
\draw[dashed] (1.5,0.5) -- (1.5, -1.2);
\draw[dashed] (7,2.5) -- (7, -1.2);
% \draw[dashed] (4,2.5) -- (4, -1.5);
\draw[dashed] (14,2.5) -- (14, -1.2);

\draw[dashed] (14.3,0.5) -- (14.3, -1.5);

\node[fill=gray, anchor=south west] at (0, 2) {Subframe 1};
\node[fill=gray, anchor=south west] at (1, 1) {Subframe 1};
\node[fill=gray, anchor=south west] at (1.5, 0) {Subframe 1};

\node[fill=yellow, anchor=south west] at (8, 1) {Subframe 2};
\node[fill=yellow, anchor=south west] at (9, 2) {Subframe 2};
\node[fill=yellow, anchor=south west] at (9.8, 0) {Subframe 2};

\node[fill=yellow, anchor=south west] at (12.5, 1) {Subframe 3};
\node[fill=yellow, anchor=south west] at (13.5, 2) {Subframe 3};
\node[fill=yellow, anchor=south west] at (14.3, 0) {Subframe 3};

\node[anchor=north west] at (1.3, -1.1) {$\delta_{\operatorname{\textsc{DLprobe}}}$ is known to attacker};
% \node[anchor=north west, text width = 3cm] at (3.8, -1.5) {Without TA, uplink arrives delayed};

\node[anchor=north, text width=6cm] at (14.3, -1.5) {Time of Arrival of Subframe 3 at \uldlsniffer{} is $\operatorname{ToA} = t_3 + \delta_{\operatorname{UE}} + \delta_{\operatorname{\textsc{ULprobe}}} - \delta_{\operatorname{TA}}$};

\draw[<->] (0,1.2) -- node[above]{$\delta_{\operatorname{UE}}$} (1, 1.2);
\draw[<->] (0,0.2) -- node[above]{$\delta_{\operatorname{\textsc{DLprobe}}}$} (1.5, 0.2);

\draw[<->] (7,1.2) -- node[above]{$\delta_{\operatorname{UE}}$} (8, 1.2);

\draw[<->] (7,2.2) -- node[above]{$2\delta_{\operatorname{UE}}$} (9, 2.2);
\draw[<->] (7,0.2) -- node[above]{$\delta_{\operatorname{UE}} + \delta_{\operatorname{\textsc{ULprobe}}}$} (9.8, 0.2);

\draw[<->] (13.5,1.6) -- node[above]{$2\epsilon$} (14, 1.6);
\draw[<->] (12.5,1.6) -- node[above]{$\delta_{\operatorname{UE}}$} (13.5, 1.6);

\node[anchor=north] at (0, -1.1) {$t_1$};
\node[anchor=north] at (7, -1.1) {$t_2$};
\node[anchor=north] at (14, -1.1) {$t_3$};

\node[fill=gray, anchor=south west] at (0, 3.3) {SF};
\node[anchor=south west] at (0.6, 3.3) {Downlink Subframe};
\node[fill=yellow, anchor=south west] at (4, 3.3) {SF};
\node[anchor=south west] at (4.6, 3.24) {Uplink Subframe};
\node[anchor=south west] at (7.6, 3.3) {$t_n$};
\node[anchor=south west] at (8.2, 3.3) {Tx/Rx Time of Subframe $n$ at eNodeB};
    
\end{tikzpicture}
    \caption{Visualization of times of arrival and delays of uplink and downlink messages.}
    \label{fig:my_label}
\end{figure*}
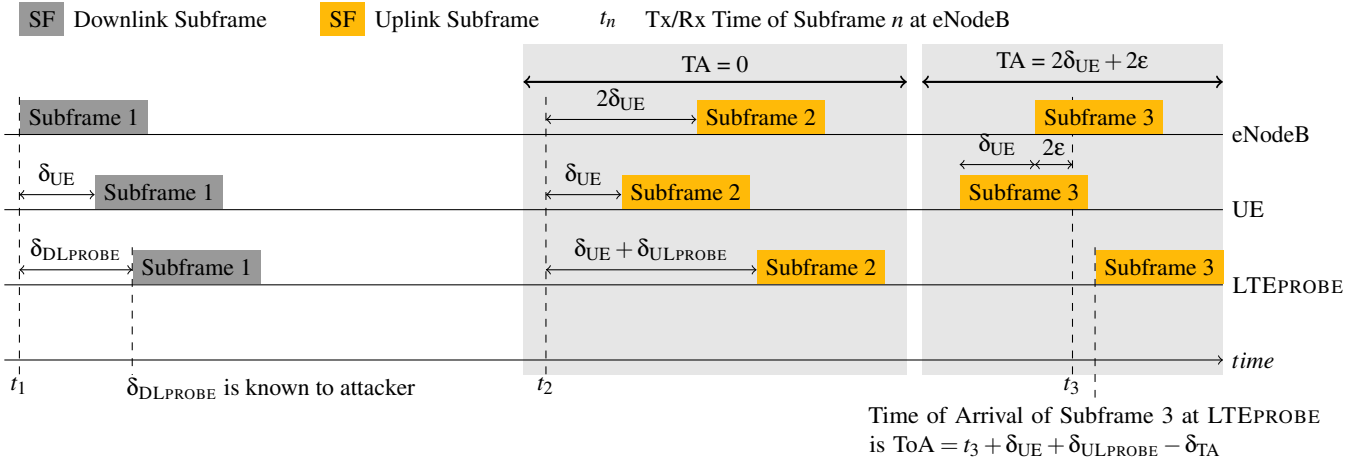

Localization attacks based on the time difference of arrival of a victim's messages constrain the victim's location to the intersection of multiple hyperbolas. The attacker can use the time difference of arrival between uplink and downlink message to define a hyperbola between the attacker and the base station. In the case of LTE, due to the systematic error introduced in the Timing Advance value, the attacker using this classical approach ends up with a 78m error. However, we show how the attacker can formulate the problem using ellipses and cancel the systematic error. We define the following variables to explain the unique localization problem in LTE:

\begin{description}[itemsep=0.1cm, parsep=0pt]
\item $t_n$ the time of the transmission of the downlink subframe $n$ by the eNodeB. Tge UE tries to send the uplink subframe $n$ such that it arrives at eNodeB at time $t_n$.

\item $\delta_{\operatorname{UE}}$ propagation delay between the eNodeB and the UE.

\item $\delta_{\operatorname{\textsc{DLprobe}}}$ propagation delay between the eNodeB and \uldlsniffer{}. We assume this value is known to the attacker since it knows the location of both the eNodeB and \uldlsniffer{}.

\item $\delta_{\operatorname{\textsc{ULprobe}}}$ propagation delay between \uldlsniffer{} and the UE.

\item $\delta_{\operatorname{TA}}$ time corresponding to the TA value received in the Timing Advance Command.

\item $\epsilon$ systematic error TA value introduces due to discretization of the propagation delay. It is the difference between the propagation delay and the TA value shown in \autoref{fig:ta_err} and its value ranges from $-0.1302\mu s$ to $0.1302\mu s$. We know that $\delta_{\operatorname{TA}} = 2\delta_{\operatorname{UE}} + 2\epsilon$.
\end{description}

The attacker measures the time of arrival of downlink and uplink messages using \uldlsniffer{} with subsample precision. The attacker uses the reference signals sent with each transmission for the timing estimation. Therefore, the attacker can collect independent measurements for each transmission and use a rolling average to smooth out any inconsistencies.

\paragraph{Downlink Message.} 

\begin{center}
    \begin{tabular}{@{}ll@{}}
     Tx at eNodeB & $t_n$ \\ \midrule
     Rx at UE & $t_n + \delta_{\operatorname{UE}}$\\
     Rx at \uldlsniffer{} & $t_n + \delta_{\operatorname{\textsc{DLprobe}}}$ \\
\end{tabular}
\end{center}

Since $\delta_{\operatorname{\textsc{DLprobe}}}$ is known to the attacker, it can compute $t_n$ from the reception time of the downlink message. The attacker can infer the times of transmission of the subsequent subframes as $t_{n+k} = t_n + k$, since the subframe length is 1ms.

\paragraph{Uplink Message without TA Command.} 

\begin{center}
    \begin{tabular}{@{}ll@{}}
     Tx at UE & $t_n + \delta_{\operatorname{UE}}$ \\ \midrule
     Rx at eNodeB & $t_n + 2\delta_{\operatorname{UE}}$\\
     Rx at \uldlsniffer{} & $t_n + \delta_{\operatorname{UE}} + \delta_{\operatorname{\textsc{ULprobe}}}$ \\
\end{tabular}
\end{center}

The UE receives all the downlink messages delayed with $\delta_{\operatorname{UE}}$, therefore its synchronization is shifted as explained in \autoref{sec:synchronization}. It will transmit uplink messages at time $t_n + \delta_{\operatorname{UE}}$ instead of $t_n$.
The Attacker computes $t_n$ from the downlink message. It can measure $\delta_{\operatorname{UE}} + \delta_{\operatorname{\textsc{ULprobe}}}$ from the reception time of the uplink message by subtracting $t_n$.

\paragraph{Uplink Message with TA Command.} 

\begin{center}
    \begin{tabular}{@{}ll@{}}
     Tx at UE & $t_n + \delta_{\operatorname{UE}} - \delta_{\operatorname{TA}} = t_n - \delta_{\operatorname{UE}} - 2\epsilon$ \\ \midrule
     Rx at eNodeB & $t_n + 2\delta_{\operatorname{UE}} - \delta_{\operatorname{TA}} = t_n - 2\epsilon$\\
     Rx at \uldlsniffer{} & $t_n + \delta_{\operatorname{UE}} + \delta_{\operatorname{\textsc{ULprobe}}} - \delta_{\operatorname{TA}} =$\\
     & $ = t_n - \delta_{\operatorname{UE}} + \delta_{\operatorname{\textsc{ULprobe}}} - 2\epsilon$ \\
\end{tabular}
\end{center}

The Attacker can no longer precisely compute $\delta_{\operatorname{UE}} + \delta_{\operatorname{\textsc{ULprobe}}}$ by subtracting $t_n$ because of the error $2\epsilon$ which can range from $-0.2604\mu s$ up to $0.2604\mu s$.

\subsection{Localization}

In the previous two subsections, we saw two sets of information that the attacker can use to localize a victim: Timing Advance Command sent by the base station on MAC layer and the times of arrival of uplink and downlink messages at \uldlsniffer{}. \autoref{fig:localization_attack_single} visualizes the attack and possible locations of the victim's phone in the environment.

The simple localization attack works by sniffing TA Commands since they are transmitted unciphered on the MAC layer of LTE protocol. Therefore, TA Command can be recorded by our \dlsniffer{}. Because of the coarse granularity due to discretization of the TA value, TA Command localization constricts possible location to a ring around the downlink sniffer with a width of 78m (yellow ring in  \autoref{fig:localization_attack_single}).

We saw in \autoref{sec:timings} how the attacker learns times of transmission of subframes $t_n$ from the time of arrival of downlink messages. When \uldlsniffer{} receives a Downlink Control Information with scheduling for uplink transmission of the victim, it decodes the uplink message and measures its time of arrival.
The time of arrival of the uplink message to \uldlsniffer{} is:
$$
\operatorname{ToA} = t_n - \delta_{\operatorname{UE}} + \delta_{\operatorname{\textsc{ULprobe}}} - 2\epsilon
$$
By subtracting the subframe transmission time $t_n$, the attacker gets a time difference of arrivals of the uplink and the downlink message. The attacker is able to then define a hyperbola of possible locations with an error $2\epsilon$. In our approach we instead subtract the subframe time $t_n$ and add the value leaked from the TA Command to learn the sum of distances:
$$
\delta_{\operatorname{UE}} + \delta_{\operatorname{\textsc{ULprobe}}} = \operatorname{ToA} - t_n + \delta_{\operatorname{TA}}
$$
Therefore, we are able to completely cancel out the systematic error $\epsilon$ from the equation. The measured sum of the two propagation delays $\delta_{\operatorname{UE}}$ and $\delta_{\operatorname{\textsc{ULprobe}}}$ constraints a set of possible locations of the victim's UE as:
$$
d_{\operatorname{UE}} + d_{\operatorname{\textsc{ULprobe}}} = c \times (\delta_{\operatorname{UE}} + \delta_{\operatorname{\textsc{ULprobe}}})
$$
, where $d_{\operatorname{UE}}$ is the distance between UE and eNodeB, $d_{\operatorname{\textsc{ULprobe}}}$ is the distance between UE and \uldlsniffer{}, and $c$ is the speed of light in the air. This constraint defines an ellipsis with two focal points: \uldlsniffer{} and the base station.

The location is now constricted to the intersection of a ring and an ellipsis shown in \autoref{fig:localization_attack_single}. Using just one sniffer, the attacker gets two narrow location areas. 

The attacker can significantly improve the precision of TA Attack by employing multiple \uldlsniffer{}s in different locations. The final UE location lies at the intersection of multiple precise ellipses. However, it introduces extra complexity and increases the cost of the attack.

\subsection{Passive Fingerprinting Attack}
\label{sec:fingerprinting}
\paragraph{Hardware Error.}

There are four hardware devices in the system: eNodeB, \dlsniffer{}, \ulsniffer{}, and the victim's UE. All four add a slight timing error due to the circuit design, length of the cables, antennas, etc. We assume the hardware error is constant for the specific model of the device. We have not observed the error changing during the experimental evaluation of the attack in \autoref{sec:results}. Software-defined radios in \uldlsniffer{} are chosen by the attacker and the base station's hardware is selected by the operator but visible to the outside world. The only device the attacker cannot foresee in the system is the victim's UE. However, the attacker can build a database with various phones and corresponding hardware errors. If it can then identify the phone type of the victim, it can look up the corresponding hardware error.

%-------------------------------------------------------------------------------
\paragraph{Passive Fingerprinting.}
%-------------------------------------------------------------------------------

To learn the hardware error introduced by the phone model, we modify and extend the attack by Shaik et al. \cite{shaik_new_2019}. This attack analyzes the uplink traffic and classifies the baseband modem of the phones connected to the cell. A baseband modem is the chip responsible for mobile network communication. The attack in \cite{shaik_new_2019} uses a relay base station to decode uplink information; therefore, this is an active attack. We instead use \uldlsniffer{} to receive uplink information. Our improvement makes the attack entirely passive, and we show how it can be used for modem and phone type fingerprinting using the decision tree model.

As a feature vector used in fingerprinting, we use the core capabilities sent in plain-text with the Attach Request. Each phone has different capabilities implemented; therefore, these messages differ significantly.

\begin{figure}
     \centering
        \scalebox{0.67}{\input{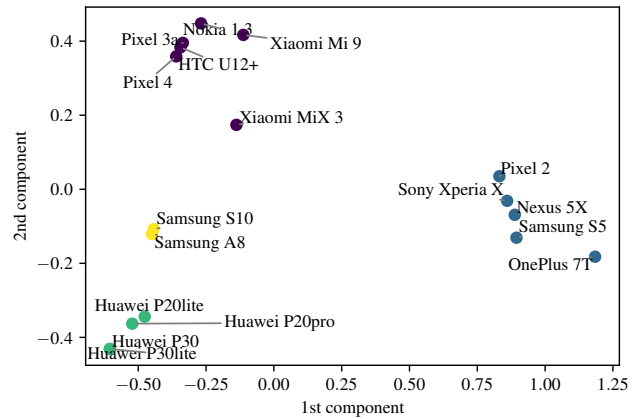}}
        \caption{First two components of PCA decomposition of the feature vector.}
        \label{fig:pca_fingerprint}
\end{figure}

\autoref{fig:pca_fingerprint} shows a PCA decomposition of the feature vector for all the tested phones, excluding iPhones. We can see that phones with the same modem manufacturer have similar core capabilities (see \autoref{tab:phones} for list of phones and their corresponding modem). We do not include iPhones in the visualisation for clarity reasons, as iPhones are clustered together far away from other phones. We can see four clusters in \autoref{fig:pca_fingerprint}. Green are the phones with Huawei modem, yellow with Samsung modem, and the blue and purple clusters are Qualcomm phones. Blue phones are older models of phones, whereas purple ones correspond to recent models. The only exception is the OnePlus 7T which was clustered with old Qualcomm phone models. Four models of phones pictured in the \autoref{fig:pca_fingerprint} have the same modem: Xiaomi Mi9, Xiaomi MiX 3 Google Pixel 4, and OnePlus 7T. OnePlus 7T is an outlier; however, the other three phones still do not have the same feature vector as they are only clustered closely together. Therefore, the capability object depends both on the modem and phone model. Thus, the attacker can learn the exact fingerprint of each phone model.

\begin{figure}
    \centering
    \begin{tikzpicture}[remember picture, 
    font=\small\ttfamily]
    
\definecolor{vlgray}{rgb}{0.9, 0.9, 0.9}
\definecolor{lgray}{rgb}{0.8, 0.8, 0.8}
\definecolor{gray}{rgb}{0.6, 0.6, 0.6}
\definecolor{dgray}{rgb}{0.5, 0.5, 0.5}
\definecolor{vdgray}{rgb}{0.2, 0.2, 0.2}

\setlength{\unitlen}{0.7cm}
\setlength{\unitlenbox}{0.7cm}

%Place a `comment` node with all text inside a tabular  
\node(comments) {\normalfont \begin{tabular}{l@{ }l}
        \subnode{A3}& \\[3mm]
        \subnode{B3}& \\[3mm]
        \subnode{C3}& \\[3mm]
        \subnode{D3}& \\[3mm]
        \subnode{E3}& \\
        \end{tabular}};

% Place `server` and `client` nodes
\node[above left=5mm and 0mm of comments, anchor=center, text width=1cm, align=center] (ue) {\normalfont \textbf{Mobile\\ Phone}};
\node[left=40mm of ue, red, anchor=center, text width=2cm, align=center] (sniffer) {\normalfont \textbf{\uldlsniffer{}\\ w/ AdaptOver}}; 
\node[left=10mm of sniffer, anchor=center, text width=1cm, align=center] (enb) {\normalfont \textbf{Base\\ Station}}; 
% \node[above=-1mm of enb] {\normalfont \textbf{Base Station}}; 
\enbnowavesnotext{($(enb) + (0.6cm, 0.5cm)$)};
\sniffer{($(sniffer) + (0, 0.5cm)$)}
\uenotext{($(ue) + (-0.35cm, 0.5cm)$)}

\definecolor{vlgray}{rgb}{0.9, 0.9, 0.9}
\definecolor{lgray}{rgb}{0.8, 0.8, 0.8}

% \filldraw [fill=vlgray, draw=vlgray] ($ (D2-|ue2) + (-4mm, 2mm) $) rectangle ($ (F2-|epc2) + (2mm, 6mm) $);
% \filldraw [fill=vlgray, draw=vlgray] ($ (H2-|ue2) + (-4mm, 4mm) $) rectangle ($ (comments2.south-|epc2) + (2mm, 0) $);
% \filldraw [fill=lgray, draw=lgray] ($ (H2-|ue2) + (-2mm, 2mm) $) rectangle ($ (comments2.south-|enb2) + (2mm, 0) $);

%Draw vertical lines
\draw [thick]($(ue)+(0,-0.5cm)$) -- (ue|-comments.south);
\draw [thick]($(enb)+(0,-0.5cm)$) -- (enb|-comments.south);
\draw [red,thick, dotted] ($(sniffer)+(0,-0.5cm)$) -- (sniffer|-comments.south);

%Draw message interchanges
\draw[->,thick] ($(A3-|ue) + (-1mm, 0)$) -- node[below]{RRC Connection Request} (A3-|sniffer) -- ($(A3-|enb) + (1mm, 0)$) ;
\draw[<-,thick] ($(B3-|ue) + (-1mm, 0)$) -- node[below]{RRC Connection Setup} (B3-|sniffer) -- ($(B3-|enb) + (1mm, 0)$) ;
\draw[->,thick] ($(C3-|ue) + (-1mm, 0)$) -- node[below]{Attach/Service Request} (C3-|sniffer) -- ($(C3-|enb) + (1mm, 0)$) ;
\draw[<-] (D3-|sniffer) -- ($(D3-|enb) + (1mm, 0)$);
\draw[<-,very thick,red] ($(D3-|ue) + (-1mm, 0)$) -- node[below]{Identity Request} (D3-|sniffer);
\draw[->,thick] ($(E3-|ue) + (-1mm, 0)$) -- node[below]{Identity Response} (E3-|sniffer) -- ($(E3-|enb) + (1mm, 0)$);

\node[red, fill=white] at ($(A3 -| sniffer)$){\faEye};
\node[red, fill=white] at ($(D3 -| sniffer)$){\faBullhorn};
\node[red, fill=white] at ($(E3 -| sniffer)$){\faEye};

\end{tikzpicture}
    \caption{The attacker sniffs a Connection Request containing the UE's TMSI. After receiving a Connection Setup, the attacker overshadows the message sent by the base station with an Identity Request message. The attacker then sniffs the Identity Response from the UE and learns its IMSI. The attacker is able to link the temporary identifier TMSI to the unique persistent IMSI.}
    \label{fig:adaptover}
\end{figure}
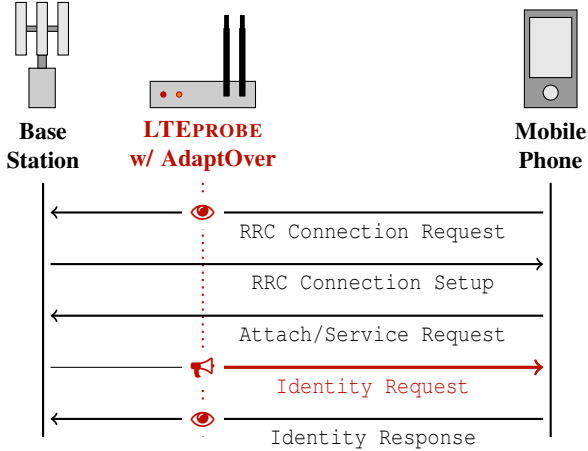

%-------------------------------------------------------------------------------
\section{IMSI Extractor}
\label{sec:identification}
%-------------------------------------------------------------------------------

To associate UEs with a unique key and therefore facilitate their tracing, we propose a new identification attack based on message overshadowing and \uldlsniffer{}. For message overshadowing, we use AdaptOver~\cite{erni_adaptover_2021}, a recently proposed LTE overshadowing attack. In AdaptOver, the attacker sends a message perfectly aligned with the base station's message timing and frequency, but with up to 3dB higher power, thus replacing the original with the attacker's message. To the UE, the attacker's messages are indistinguishable from legitimate messages. 

In this section, we show that by combining sniffing on the uplink and AdaptOver injecting just one adversarial message, we can get the UE to leak the IMSI. Since each SIM card has a unique, persistent IMSI number, the attacker perfectly distinguishes the victim with this attack. Even though the attack is active, the attacker can choose the granularity of when it wants to perform the attack as well as only target specific UEs. Our attack is triggered when the eNodeB sends a RRC Connection Setup, as pictured in \autoref{fig:adaptover}. This happens when the UE goes from an off or idle state to a connected state (e.g., the phone receives a paging message or needs to transmit data).

\paragraph{Identification of a Victim.}
As shown in the \autoref{fig:adaptover}, the UE sends an initial RRC Connection Request containing its TMSI number. However, because the LTE network can change this identifier at any time, the attacker does not have any assurance about UE's long-term identity. Instead, the IMSI number satisfies this, but the UE does not transmit IMSI in plain-text in the usual behavior of the protocol. Nevertheless, the LTE protocol allows the core of the network to request the IMSI number at any time (e.g., when the network loses the TMSI number) by sending an Identity Request.

\subsection{Overshadowing with Identity Request}

Specified in \cite{3gpp.24.301}, an Identity Request for the IMSI number can be sent by the eNodeB without any integrity protection before the security context is created. Since the security context is not set up before the Service or Attach Request, the attacker can inject an Identity Request as a response to those requests. The UE will respond to the Identity Request with an Identity Response message containing its unique IMSI number, which \uldlsniffer{} receives. \autoref{fig:adaptover} shows the message exchange. Even though the legitimate base station proceeds with a connection procedure, AdaptOver sends a message with a higher power, overshadowing it. Thus, the UE only decodes the Identity Request that is sent by the attacker. The base station does not receive the Identity Response sent by the UE, because AdaptOver also modifies the uplink allocation during the attack. Overall the attack requires a limited number of transmissions by the attacker, with only a slightly higher power than the base station. 

It is essential to point out that using Identity Request is just one concrete approach to how IMSI Extractor can operate. However, the attacker is not constrained by this and can create other protocol compliant communication traces, which trigger IMSI transmission in plain-text by the UE (e.g., Service Reject with cause 9, ``UE identity cannot be derived by the network").

We present the first attack that combines the overshadowing attack with an uplink sniffing to violate user privacy. Earlier overshadowing attacks like SigOver~\cite{yang_hiding_2019} and AdaptOver~\cite{erni_adaptover_2021} focused on denial of service.  

\paragraph{Stealthiness of Our Attack.}
To a UE, the message exchange with a spoofed Identity Request looks benign. According to the LTE specification \cite{3gpp.24.301}, a network can start an identification procedure at any time, even right after it received an Attach Request or Service Request. Therefore, from the protocol-level point of view at the UE, our attack does not raise any alarms. The base station also does not notice any problems. From the perspective of the eNodeB, the connection with the UE halted (e.g., due to bad reception at the UE). For both the UE and the base station, the traces generated by the attacker's messages are therefore compliant with the protocol. 

Current detection mechanisms against IMSI Catchers work by detecting fake base stations~\cite{nakarmi_prajwol_kumar_3gpp_2019, dabrowski_imsi-catch_2014, quintin_detecting_2020, li_fbs-radar_2017,borgaonkar_white-stingray_2017}. These frameworks either work by comparing open-sourced locations of base stations to measured reports by users or special devices, or by detecting anomalies in the behavior of base stations by UEs. In case of our attack, these techniques do not work since a UE connects to a real base station. Therefore, to UEs, the behavior and location of the cell are legitimate. As proposed in~\cite{echeverria_phoenix_2021}, a signature based anomaly detector with a signature: ``if Identity Request, then attack'', is successful in the detection of our attack. However, because Identity Requests are also sent during a legitimate protocol flow, such a solution will inherently report false positives during legitimate identification procedures. Moreover, the attacker is not constrained to sending Identity Requests to perform IMSI Extractor, as mentioned above.

As explained in~\cite{borgaonkar_white-stingray_2017, dabrowski_imsi-catch_2014}, other features (e.g., number of neighbouring cells) are considered in most of the anomaly-based IMSI Catcher catching apps. Evaluating them, our catcher is not classified as an IMSI Catcher. Therefore, we consider our attack to be stealthy, at least with respect to existing deployed and proposed techniques.

\begin{figure}
    \centering
    \begin{tikzpicture}[remember picture, 
    font=\small\ttfamily]
    
\definecolor{vlgray}{rgb}{0.9, 0.9, 0.9}
\definecolor{lgray}{rgb}{0.8, 0.8, 0.8}
\definecolor{gray}{rgb}{0.6, 0.6, 0.6}
\definecolor{dgray}{rgb}{0.5, 0.5, 0.5}
\definecolor{vdgray}{rgb}{0.2, 0.2, 0.2}

\setlength{\unitlen}{0.7cm}
\setlength{\unitlenbox}{0.7cm}

\clip (-1,0) rectangle (7,8);

\fill[vlgray] (-1,0) rectangle (7,5);

\fill [green] plot [smooth] coordinates {(1.6,2.6) (2.2,2.6) (3,2.3) (3.8,2.6) (4.6,2.5) (4.8,1.3) (5.1,-0.2) (1.8,0.3) (1.6,2.6)};
\fill [blue] plot [smooth cycle] coordinates {(2.5, 2) (3.5, 2) (3.7, 1.3) (2.7,1)};

\fill[vlgray] (5.1,0) rectangle (7,5);

% roads

\fill[white] (-1.5,2.5) rectangle (7.5,2.7);

\fill[white] (1.5,0) rectangle (1.7,5);

\fill[white] (4.9,0) rectangle (5.1,5);
\fill[white] (-0.1,0) rectangle (0,5);

\fill[white] (3.5,2.6) rectangle (3.6,5);

\fill[white] (-1.5,0.75) rectangle (1.6,0.8);

\fill[white] (-1,3.7) rectangle (7,3.9);

% buildings
\fill[lgray] (1.8, 2.8) rectangle (2.2, 3.6);
\fill[lgray] (2.3, 2.8) rectangle (3.3, 3.1);
\fill[lgray] (2.3, 3.3) rectangle (3, 3.6);
\fill[lgray] (3.1, 3.2) rectangle (3.4, 3.6);

\fill[lgray] (1.8, 4) rectangle (2.8, 4.4);
\fill[lgray] (2.9, 4) rectangle (3.4, 4.9);
\fill[lgray] (1.8, 4.5) rectangle (2.3, 4.9);

\fill[lgray] (0.1, 0.4) rectangle (0.9, 0.7);
\fill[lgray] (1, 0.1) rectangle (1.4, 0.65);

\fill[lgray] (0.1, 0.9) rectangle (1.4, 1.7);
\fill[lgray] (0.1, 1.8) rectangle (0.4, 2.4);
\fill[lgray] (0.5, 1.9) rectangle (0.9, 2.4);
\fill[lgray] (1, 1.8) rectangle (1.4, 2.4);

\fill[lgray] (-0.7, 0.3) rectangle (-0.2, 0.65);

\fill[lgray] (-0.9, 0.85) rectangle (-0.2, 1.25);
\fill[lgray] (-0.9, 1.35) rectangle (-0.2, 1.75);
\fill[lgray] (-0.9, 1.85) rectangle (-0.2, 2.4);

\fill[lgray] (-0.6, 2.8) rectangle (-0.2, 3.6);

\fill[lgray] (-0.9, 4) rectangle (-0.2, 4.5);

\fill[lgray] (0.1, 4) rectangle (0.8, 4.4);
\fill[lgray] (0.9, 4) rectangle (1.4, 4.35);
\fill[lgray] (0.9, 4.4) rectangle (1.4, 4.8);

\fill[lgray] (0.1, 3.2) rectangle (0.8, 3.6);
\fill[lgray] (0.1, 2.8) rectangle (0.7, 3.1);

\fill[lgray] (4.3, 2.8) rectangle (4.8, 3.1);
\fill[lgray] (4.4, 3.2) rectangle (4.8, 3.6);
\fill[lgray] (3.7, 2.8) rectangle (4.2, 3.6);

\fill[lgray] (3.7, 4.5) rectangle (4.2, 4.9);

\fill[lgray] (5.2, 4.5) rectangle (5.8, 4.9);
\fill[lgray] (5.2, 4) rectangle (6.1, 4.4);
\fill[lgray] (5.9, 4.5) rectangle (6.9, 4.9);
\fill[lgray] (6.5, 4) rectangle (6.9, 4.9);

\fill[lgray] (5.2, 3.3) rectangle (6.3, 3.6);
\fill[lgray] (5.2, 2.8) rectangle (5.7, 3.2);
\fill[lgray] (5.8, 2.8) rectangle (6.2, 3.2);
\fill[lgray] (6.3, 2.8) rectangle (6.9, 3.2);

\fill[lgray] (5.2, 2.4) rectangle (5.7, 1.5);
\fill[lgray] (5.2, 2.4) rectangle (6, 1.9);
\fill[lgray] (6.1, 1.4) rectangle (6.8, 2.3);
\fill[lgray] (5.2, 0.6) rectangle (5.6, 1.4);

% UE

% \fill (3.5, 4.4) circle (0.05);
% \fill (3.55, 3.4) circle (0.05);

% \fill (3.75, 1.2) circle (0.05);
% \fill (3.95, 1.35) circle (0.05);
% \fill (3.35, 0.9) circle (0.05);
% \fill (3.55, 1) circle (0.05);
% \fill (3.65, 1.25) circle (0.05);
% \fill (3.55, 1.1) circle (0.05);

\scalebox{0.5}{\uenotext{(7,2.2)}}

\scalebox{0.5}{\uenotext{(6.6,1.8)}}
\scalebox{0.5}{\uenotext{(8,1.6)}}
\scalebox{0.5}{\uenotext{(7.3,1)}}
\scalebox{0.5}{\uenotext{(5,1)}}
\scalebox{0.5}{\uenotextyellow{(7.4,2.4)}}

\scalebox{0.5}{\uenotext{(12,4.8)}}

\scalebox{0.5}{\uenotext{(3,8)}}
\scalebox{0.5}{\uenotext{(0,4.5)}}

\scalebox{0.5}{\uenotext{(-0.4,1.5)}}

\scalebox{0.5}{\enbnotext{(3,5.8)}}
\scalebox{0.5}{\sniffer{(8.5,8)}}
\scalebox{0.5}{\sniffer{(11.5,0.5)}}

\draw[->,red,very thick] (4,2) -- node[text width = 2cm,right] {\textbf{Identity Response}} (4.2,3.8);
\draw[<-,red,very thick] (3.8,2) -- node[text width = 1.4cm,left] {\textbf{Identity Request}} (4,3.8);

\node[align = center] at (3, 7.5) {\normalfont \Large \textsc{LTrack}};

\node[align = center] at (1, 6) {\normalfont \large \textcolor{red}{IMSI Extractor}};
\node[align = center] at (5, 6) {\normalfont \large \textcolor{red}{Passive Localization}};
\node[align = center] at (5, 5.5) {\normalfont \uldlsniffer{}};
\node[align = center] at (1, 5.5) {\normalfont AdaptOver + \uldlsniffer{}};

\draw[very thick] (1,6.75) -- (5,6.75);
\draw[very thick] (1,6.75) -- (1,6.3);
\draw[very thick] (3,6.75) -- (3,7.2);
\draw[very thick] (5,6.75) -- (5,6.3);

\end{tikzpicture}
    \caption{Visualization of \textsc{LTrack}, tracking attack based on passive localization and IMSI Extractor. When the attacker loses track of a victim in a natural mix-zone, the attacker uses the IMSI Extractor to distinguish the victim from other UEs in the area.}
    \label{fig:map}
\end{figure}
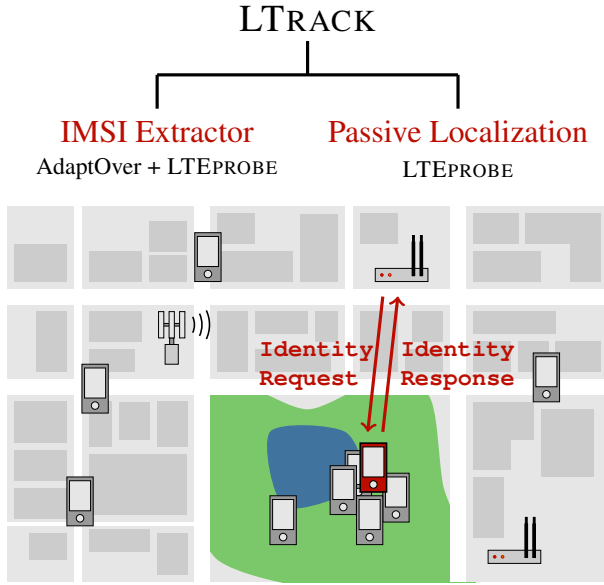

%-------------------------------------------------------------------------------
\section{\textsc{LTrack}}
\label{sec:tracking}
%-------------------------------------------------------------------------------

In this section, we discuss how the techniques that we introduced in \autoref{sec:localization} and \autoref{sec:identification} can be put together to support large-scale tracking attacks, similar to those described e.g., in~\cite{shokri_quantifying_2011}. The goal of such a tracking attacks is to obtain traces of all users while staying as stealthy as possible. 

\autoref{fig:map} visualizes a city setting where the attacker tries to localize users. The attacker uses the passive localization attack to locate individual users during their connections to base stations. However, without identification, all the UEs look the same, and after each reconnection, UEs might anonymize themselves with a new TMSI. If the user moves along less frequented areas when the TMSI gets updated, the attacker could still link the two temporary identifiers based on their locations. However, as an UE enters an area with many other UEs, this area will act as a natural mix zone. \textsc{LTrack} solves this problem by using a combination of passive tracking and IMSI Extractor, allowing the attacker to distinguish UEs.

In order to launch our attack on a large-scale, the attacker needs to deploy at least one, preferably two, \uldlsniffer{}s for each base station the attacker decides to monitor. \uldlsniffer{}s are placed away from the base stations such that the attacker can perform the localization attack pictured in \autoref{fig:localization_attack_single}. We do not put any restrictions on the attacker in terms of available funds or access to the locations. The attacker can place its devices at high vantage points (e.g., skyscrapers or communication towers). Building such a network of devices is feasible. Competitor service providers often already have devices (base stations) at preferred locations, which they can transform into sniffers.
Our attack works in the following four stages:

\paragraph{(i) Communication Recording.} The attacker uses \uldlsniffer{}s to passively record all the traffic on the set of base stations it monitors. All the uplink and downlink communication with corresponding arrival times from all \uldlsniffer{}s is stored in the attacker's database. All messages during a UE's connection to the eNodeB are addressed on the physical layer by a unique RNTI value as explained in \autoref{sec:sniffer}, linking the messages together. Moreover, each connection of a UE to the network starts with an RRC Connection Request containing the TMSI of the user. The attacker stores a list of TMSIs observed during the attack's execution and links them to the corresponding connections. To link connections of the same user, whose TMSI changed during the execution, the attacker runs IMSI Extractor, described below. 

\paragraph{(ii) IMSI Extractor.} IMSI Extractor extracts and stores TMSI-IMSI pairs of the users, linking observed communication to the unique, persistent identifier of the UE (IMSI) as explained in \autoref{sec:identification}. Once \uldlsniffer{} registers RRC Connection Request, the attacker checks whether it already knows the corresponding IMSI to the enclosed TMSI inside the RRC Connection Request. If the TMSI-IMSI pair exists in the database, the attacker does not engage, passively records the communication, and links the communication to the stored pair. However, if the TMSI has not been seen before, it runs the IMSI Extractor to learn the IMSI number and stores the new TMSI-IMSI pair in the database.

\paragraph{(iii) Passive Localization.} Finally, the attacker has recordings of all the users at multiple base stations under different TMSI-IMSI pairs. The attacker uses recorded data to get each UE uplink message's time of arrival and Timing Advance Commands sent by the base station. As shown in \autoref{fig:localization_attack_single}, each uplink message measurement constrains possible locations of the user. 

Moreover, the attacker runs a passive fingerprinting attack on the saved recordings of Attach Requests to learn the phone model. With the model of the phone, the attacker can increase the precision of the localization attack. Furthermore, since the attacker stores all the recorded communication, it can retroactively compensate hardware error to increase the precision of measured times of arrival of an uplink messages for that user. Therefore, even if the user's TMSI changes, we will update it in the database during the subsequent Service/Attach Request.

We can further improve the precision of localization by choosing more likely locations, e.g., the user probably moves along the street, not through the walls of buildings. Altogether, the attacker can visualize the movement of the victim. Finally, the attacker builds a whole trace of a user's movement.

\paragraph{(iv) Special Cases.} Under certain conditions (i.e., handover), a UE stays connected to the network but changes the serving cell to a cell with a stronger signal. Then, the UE disconnects from the old cell and performs a random access procedure with the new cell. Since it is still connected to the network, there is no need for a Service Request message. Thus, the attacker can observe new random access without a Service Request, and it can match it to a connection that halted at a neighboring cell. If the localization attack is in place, the attacker can improve the matching between the old and new connections based on the location of the UE.

Even if the attacker loses track of a UE, the attacker observes the UE again during the next Service Request it performs. For example, for an inactivity timer of 10 seconds, on average, a UE connects to the network more than once per minute under background traffic (i.e., a user does not actively use a phone) \cite{3gpp.36.822}, which is a usual scenario during a movement of a person. The attacker can also force a reconnection using a paging message or a call.

\paragraph{}

In this work we do not address user deanonymization. Research in this area is already quite extensive and the attacker can use multiple existing techniques to obtain true user identities. User traces reconstructed by \textsc{LTrack} can be used to identify users~\cite{wernke_classification_2014, krumm_survey_2009}, for example, based on transportation routines~\cite{liao_learning_2007}, mobility traces~\cite{wang_-anonymization_2018, pyrgelis_knock_2018, zang_anonymization_2011, gambs_-anonymization_2014}, home addresses~\cite{hoh_enhancing_2006, lamarca_inference_2007, tokuda_anonymity_2009}, who they meet~\cite{olteanu_quantifying_2014, srivatsa_deanonymizing_2012}, or online geo-tagged media~\cite{henne_snapme_2013}. \cite{de_montjoye_unique_2013,de_mulder_identification_2008} show that even coarse spatial and temporal traces deanonymize the users based on their unique mobility patterns. 

%-------------------------------------------------------------------------------
\section{Experimental Evaluation}
\label{sec:results}
%-------------------------------------------------------------------------------

\subsection{Experimental Setup}

%---------------------------
\begin{figure}
\begin{center}
\includegraphics[width=\linewidth]{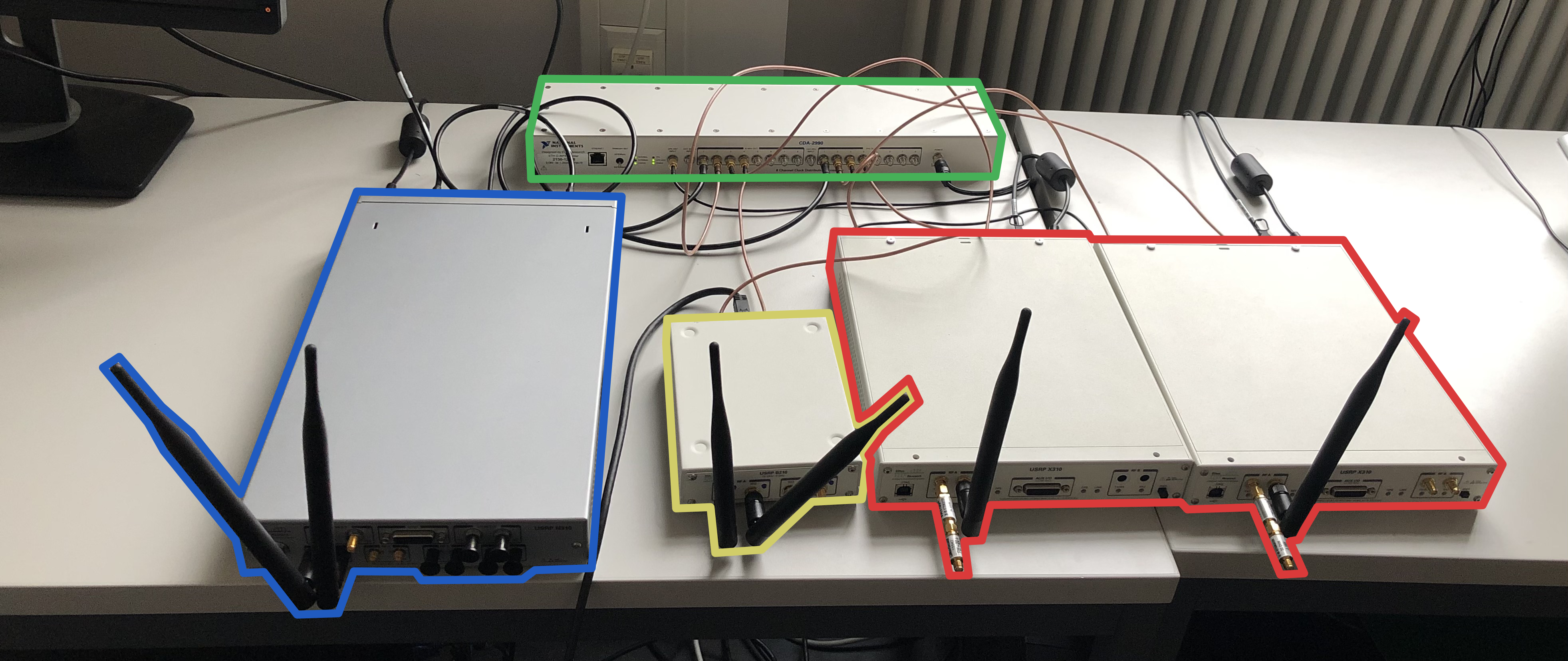}
\end{center}
\caption{\label{fig:setup} Our setup used for the evaluation of the passive localization and the IMSI Extractor. }
\end{figure}
%% %---------------------------

For the experimental evaluation of our attack we used the setup pictured in \autoref{fig:setup}. It consists of:

\begin{description}[itemsep=0.1cm, parsep=0pt]
\item[eNodeB] running on software defined radio USRP N310, highlighted in the blue color in \autoref{fig:setup}. Alternatively, we use an entry-grade base station AMARI Callbox Mini~\cite{amarisoft_amari_nodate} for the evaluation of IMSI Extractor attack. However, due to its lower grade clock, the timing is inaccurate. Thus, we do not use it for the localization attack, where the accuracy is necessary. 
\item[\uldlsniffer{}] running on two USRP X310 SDRs, highlighted in the red color in \autoref{fig:setup}. One X310 is used as a \dlsniffer{} and the other as \ulsniffer{}. There is no antenna connected to the Tx port of the radios confirming it is a passive device. Both devices are connected to the Octoclock to share the same clock.
\item[Octoclock] model CDA-2990, highlighted in the green color distributes the same clocking signal to all connected devices. It takes the GPS signal as input. All connected devices have the same sense of time. The two sniffing USRPs are always connected to Octoclock.
\item[AdaptOver] running on software defined radio USRP B210, highlighted with in the yellow color in \autoref{fig:setup}. 
\item[UE] During the experimental evaluation, we use multiple mobile phones as UEs. The full list of UEs is recorded in \autoref{tab:phones} and \autoref{tab:identification_phones} in the Appendix.

\end{description}

\subsection{Passive Localization Attack}

\begin{figure*}
      \centering
        \input{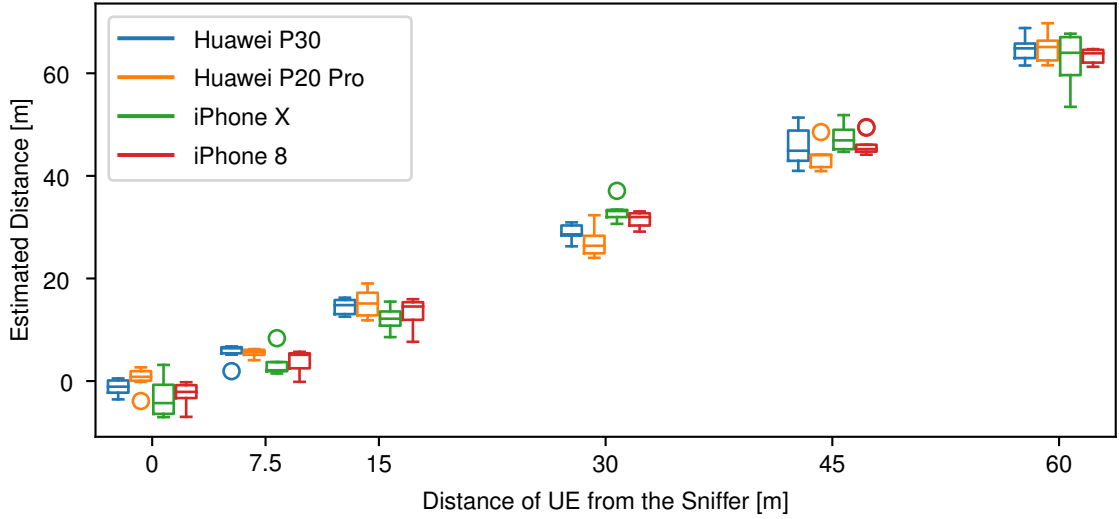}
        \caption{Distance measurements for four different phone at six distances.}
        \label{fig:distance}
\end{figure*}

For the experimental evaluation of our localization attack, we collocated the eNodeB and \uldlsniffer{}, and we varied the location of UEs. Instead of the location, we estimated the distance of the \uldlsniffer{} from a UE. In our experiment, we learn the measurement error of \uldlsniffer{}, which we can use to quantify the localization error under various dilutions of precision. Since eNodeB and \uldlsniffer{} are at the same location, the distance of the UE from \uldlsniffer{} is:

$$
d_{\operatorname{\textsc{ULprobe}}} = c \times (\delta_{\operatorname{UE}} + \delta_{\operatorname{\textsc{ULprobe}}}) / 2
$$

We conducted the experiment with five different UEs: USRP B210 with srsUE, Huawei P20 Pro, Huawei P30, iPhone X, and iPhone 8. We positioned the UE in line-of-sight at six different distances in a long corridor indoors: 0$m$, 7.5$m$, 15$m$, 30$m$, 45$m$, and 60$m$. For each distance and UE, we reconnected six times to measure the distance over multiple connections. For each distance measurement and UE, we restarted \uldlsniffer{} at least once to reset the synchronization errors. 

The accuracy of the internal clock without a GPS is $\pm$2.5ppm and $\pm$0.1ppm for the USRP X310 and USRP N310 respectively. This accuracy improves to $\pm$0.01ppb for both types of devices when the GPS lock is acquired. Since we could not acquire a GPS lock in our environment, we instead had both the eNodeB and the \uldlsniffer{} connected to Octoclock, which we use as a proxy to the GPS acquisition in a real world scenario. Assuming Octoclock provides perfect accuracy (0ppb), using a GPS locked clock instead of Octoclock introduces a synchronization error of merely $\pm$0.02ppb. This error would translate to an error of 40nm while performing distance measurement over 1 km. We therefore consider the improvement of using Octoclock negligible. We did not use Octoclock to synchronize the attacker's devices and the eNodeB.

In our experiment, one data point corresponds to the median distance measurements during one connection of the UE to our eNodeB. We do not consider connections for which we have less than ten measurements. For each UE, there is a constant hardware error that comes from properties of UE modem, \uldlsniffer{} radios, and eNodeB radio. We estimate constant hardware error as a mean difference between estimated distances and actual distances. Before plotting, we remove hardware error from these distance estimations. Finally, we visualize data points with boxplots for each UE in \autoref{fig:distance}. 

In \autoref{tab:phones}, we quantify the constant hardware error for all the test phones as well. We estimate the hardware error with a single distance measurement at 0$m$. We observe that the hardware error is the same for all UEs with the same LTE baseband modem. Moreover, all Intel modems have the same error.

To quantify distance estimation error, we compute errors between estimated variables (with corrected constant hardware error) and actual distances. The distance estimation error can be directly translated into localization error under the ideal dilution of precision.
We observe that for all mobile phones the 90th percentile error is $\sim 6m$. Concretely, the 90th percentile of the errors are: 5.659m for Huawei P20 Pro, 5.214m for Huawei P30, 7.238m for iPhone X, and 4.672m for iPhone 8. For USRP B210 the 90th percentile is 10.474, however, the performance of B210's clock is limited without a GPS lock. Obviously, for lower percentile, the values get significantly better. Median error is $\sim 2m$ for phones and $\sim 7m$ for B210.

One of the problems we observed was an error arising from the UE not receiving the TA Command. If the UE does not receive the TA Command, the eNodeB resends it. However, \uldlsniffer{} receives it twice and applies the command again, resulting in a mismatch. Since the N310 is not a professionally graded eNodeB device, its Tx power is lower. We can expect better performance in the real world. A possible fix in the future would be to monitor ACKs sent by the UE. \uldlsniffer{} would then only apply TA Commands that the UE acknowledged. We removed connection outliers that were more than ten times the interquartile range away from the median point. Out of 186 connections, we removed 4 data points.

\subsection{IMSI Extractor}

Since IMSI Extractor is a protocol-level attack, we evaluate it using an industry-grade base station software by Amarisoft on the AMARI Callbox Mini hardware~\cite{amarisoft_amari_nodate}. The base station, the AdaptOver USRP and the Octoclock used GPS clock.

We ran the attack against 17 modern phones for both Attach Request and Service Request messages. For all 17 phones, we obtained the IMSI number as a response to the Attach Request. As a response to the Service Request, we were successful for all but one mobile phone, iPhone 7. After transmitting the Identity Response, the UEs successfully connected to the network. To the user, the attack was not noticeable. The comprehensive list of phones used in the evaluation and an example packet capture file from our attack can be found in Appendix in \autoref{tab:identification_phones} and \autoref{fig:pcap}.

Finally, we confirmed our attack and the capabilities of \uldlsniffer{} against a live network of a national operator. The setup consisted of a real-world Ericsson eNodeB, connected to the operator's production core network, with its antennas and our attacker devices installed inside a 5$\times$6m Faraday cage. Therefore, we could run tests against the same configuration as found in outside cells, without influencing real users.

%-------------------------------------------------------------------------------
\section{Countermeasures}
\label{sec:countermeasures}
%-------------------------------------------------------------------------------

As shown in~\cite{rasmussen_location_2008}, it is impossible to mitigate location leakage attacks presented in this work unless messages and their transmission/reception times are fully randomized. Due to the highly synchronized operation of LTE, these requirements are not feasible to be implemented. 

Instead, we propose a solution that only requires changes on UEs and is compatible with the current LTE protocol. In our countermeasure, UE sends the initial Random Access message with a random offset. Since UE knows the offset, it modifies the received Timing Advance Command by adding the applied random offset. The recorded Timing Advance value by \uldlsniffer{} is therefore not relevant and using it in the localization attack results in wrong location estimates. Our proposal does not mitigate the localization attack, but increases its complexity and cost. The attacker can employ more sniffers and infer the random offset UE applies.

We propose three types of countermeasures against our IMSI Extractor: 
\begin{enumerate*}[label=(\roman*)]
\item UE-based countermeasures are deployed on the UEs and work by observing Identity Requests for the IMSI number. UEs notify users about incoming Identity Requests or report to the network an unusual number of Identity Requests. Reporting to the operator requires trust in the UEs that they report the numbers honestly.
\item Network-based countermeasures use a large number of eavesdroppers in the covered area. They compare the eavesdropped Identity Requests with the ones sent by the base stations. Since the operators deploy the eavesdroppers, they have access to all the transmitted Identity Requests. Neither UE-based nor network-based countermeasures prevent IMSI Extractor but merely detect it. 
\item Finally, Protocol-based countermeasures are the most robust and work even against IMSI Extractor based on other procedures; however, they require the most extensive changes to LTE, likely unfeasible to retrofit for existing devices. In 5G, IMSI catching is no longer possible since IMSI is encrypted using the network's public key. Thus, the attacker cannot decode the IMSI.\end{enumerate*}

%-------------------------------------------------------------------------------
\section{Related Work}
\label{sec:realted_work}
%-------------------------------------------------------------------------------

The first paper to implement a downlink control channel sniffer was by Kumar et al.\cite{kumar_lte_2014}. The follow up work by Bui et al.\cite{bui_owl_2016} implements a downlink control channel sniffer with the open-source library srsLTE \cite{gomez-miguelez_srslte_2016}. 
We improve on these two papers with a downlink sniffer that decodes data channels and reconstructs higher layer datagrams. This allows us to receive TA Commands on the MAC layer or get a UE dedicated configuration for the uplink channel on the RRC layer. However, neither of these works implements an uplink sniffer functionality, which is paramount for \textsc{LTrack}.

Three commercial sniffers are available. Airscope~\cite{software_radio_systems_products_nodate} is a downlink-only sniffer, whereas Wavejudge \cite{sanjole_wavejudge_nodate} and thinkRF \cite{thinkrf_leader_nodate} cover both uplink and downlink sniffer functionality. These products are high price and closed source, so we could not compare our sniffer to these products nor use them to mount our attacks.

In terms of user tracking and localization, 
Shaik et al. \cite{shaik_practical_2017} show how an adversary may sniff on paging messages at different eNodeBs. An operator will first broadcast a paging message for a particular user from the last used eNodeB. From this, the attacker learns a coarse location of the UE.

In LTEye \cite{kumar_lte_2014}, the authors extend a synthetic aperture radar to capture the shortest and the most direct path of the radio signal from the User Equipment. The users' location is estimated at the intersection of the direct paths, estimated by multiple radars at different locations.

The closest work to ours (in the context of UE localization) is the work of~\cite{roth_location_2017}. \cite{roth_location_2017} also proposes the use of both Timing Advance Command from eNodeB and times of arrival of uplink messages to approximate the geolocation of the UE. However,~\cite{roth_location_2017} does not provide details regarding the measurement of the time of arrival from uplink messages, does not implement the attack, nor does it bind the obtained location with the UE identity as we do in this paper. In particular, in our work, we also increase the localization accuracy by fingerprinting the phone model and correcting its hardware error. \cite{roth_location_2017} opted for approximating transmission time of UE from Timing Advance Command, which introduces a significant error. We transform the geometry of the problem into an ellipse with two focal points, which cancels the large systematic error introduced by the Timing Advance Command. \cite{roth_location_2017} further highlights how their work is successful in a setting where the UE is in the vicinity of multiple eNodeBs. Having a single eNodeB close to the victim with a deployed sniffing device is sufficient in our attack.

\cite{pimentel_passive_2013} show real-world localization attack based on Timing Advance Commands against WiMax networks.
They used a commercial device, WaveJudge 4900A \cite{sanjole_wavejudge_nodate} to perform the attack. 
They improve the distance estimation of mobile phones from the base station by using time of arrival measurement. 
However, compared to that work, our time of arrival estimation offers subsample precision. \cite{pimentel_passive_2013} further didn't evaluate modern smartphones and their corresponding hardware errors.

So far, the primary tool for UE identification were IMSI Catchers, which rely on fake base stations~\cite{shaik_practical_2017, jover_lte_2016} and are therefore detectable. 
Other approaches included triggering reconnections by forcing the victim's UE to act, e.g., by sending a WhatsApp or Facebook message \cite{kohls_lost_2019, shaik_practical_2017}. When UE reconnects to the network, the attacker can infer the model and make of the device and compare it to the victim's UE~\cite{shaik_new_2019}. However, such attacks are primarily targeted against specific UEs and are not sufficient for large-scale tracking. 

%-------------------------------------------------------------------------------
\section{Conclusion}
\label{sec:conclusion}
%-------------------------------------------------------------------------------

In this work, we proposed and showed the feasibility of large-scale tracking of users in an LTE network. Furthermore, we built \uldlsniffer, a robust uplink and downlink sniffer based on components of srsLTE.
The implementation of \uldlsniffer{} is white-box and does not depend on any costly or proprietary modules other than off-the-shelf software-defined radios. Using our sniffer, we were able to devise a tracking attack that we call \textsc{LTrack}.
\textsc{LTrack} improves on the state-of-the-art by combining Timing Advance Command sniffing and measuring the times of arrival of both LTE downlink and uplink messages. \textsc{LTrack} also contains a purpose-built IMSI Catcher that does not rely on a fake base station but rather overshadows packages with surgical precision and very little energy.
This work is the first to explore UE tracking in a practical setting and with affordable hardware.

%-------------------------------------------------------------------------------
\section*{Acknowledgment}
\label{sec:acknowledgment}
%-------------------------------------------------------------------------------

This project has received funding from the Swiss National Science Foundation under NCCR Automation, grant agreement 51NF40\_180545.

%-------------------------------------------------------------------------------
\small
\bibliographystyle{plain}
\bibliography{references}

\appendix 
\section{Appendix}

\begin{figure*}
    \centering
    \includegraphics[width=0.8\textwidth]{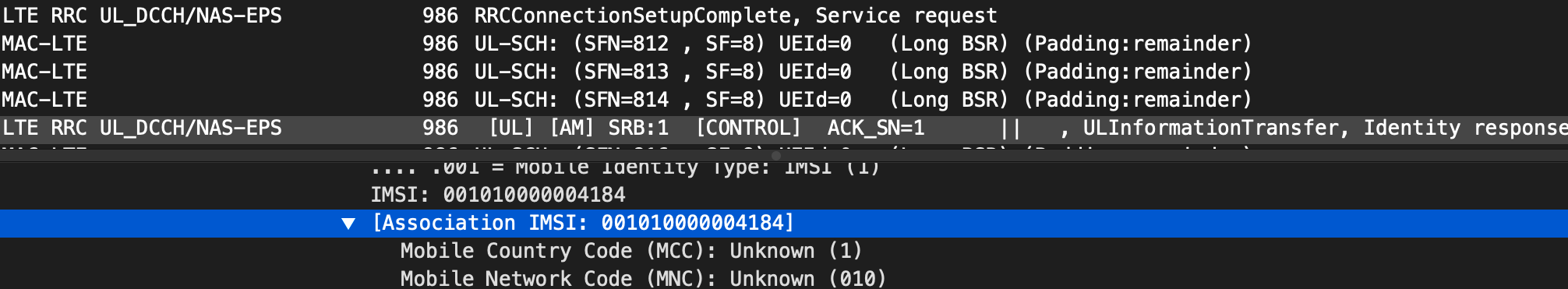}
    \caption{Packet capture file from IMSI Extractor.}
    \label{fig:pcap}
\end{figure*}

\begin{table}[h]
\smaller
\ra{1.3}
\centering
\begin{tabular}{@{}lrr@{}}\toprule
UE model & \makecell[r]{Identification \\ Attach Request} & \makecell[r]{Identification \\ Service Request} \\ \midrule
Samsung Galaxy s10 & yes & yes\\
Samsung Galaxy a8 & yes & yes\\
Huawei P20 Pro & yes & yes\\
Huawei P30 Lite & yes & yes\\
Huawei P30 & yes & yes\\
Xiaomi Mi9 & yes & yes\\
Xiaomi MiX 3 & yes & yes\\
Google Nexus 5X & yes & yes\\
Google Pixel 2 & yes & yes\\
Google Pixel 3a & yes & yes\\
HTC U12+ & yes & yes\\
OnePlus 7T & yes & yes\\
iPhone 6s & yes & yes\\
iPhone 7 & yes & no\\
iPhone 8 & yes & yes\\
iPhone X & yes & yes\\
iPhone 11 & yes & yes\\
iPhone 11 Pro & yes & yes \\\bottomrule
\end{tabular}
\caption{Mobile phones used in the IMSI Extractor experiments.}
\label{tab:identification_phones}
\end{table}

\begin{table}[h]
\smaller
\ra{1.3}
\centering
\begin{tabular}{@{}lrrr@{}}\toprule
UE model & Modem & \makecell[r]{Hardware\\Error [$m$]} & std [$m$]\\ \midrule
Samsung Galaxy s10 & Exynos 9820 & 11.29 & 7.22\\
Samsung Galaxy a8 & Exynos 7885 & -26.62 & 4.77\\
Samsung Galaxy s5 & Qcom. Gobi 4G & - & -\\
Huawei P20 Lite& Kirin 659 & -24.47 & 2.13\\
Huawei P20 Pro & Kirin 970 &  -9.34 & 2.90\\
Huawei P30 Lite& Kirin 710 & -10.27 & 0.98\\
Huawei P30 & Kirin 980 & -24.51 & 1.49\\
Xiaomi Mi9 & Qcom. X24 LTE & 10.44 & 2.20\\
Xiaomi MiX 3 & Qcom. X24 LTE & 11.57 & 1.60 \\
Nokia 1.3 & Qcom. X5 LTE & - & - \\
Sony Xperia X & Qcom. X8 LTE & -11.20 & 4.78 \\
Google Nexus 5X & Qcom. X10 LTE & 5.08 & 2.51\\
Google Pixel 2 & Qcom. X16 LTE & -13.52 & 2.32\\
Google Pixel 3a & Qcom. X12 LTE & 4.46 & 2.14\\
Google Pixel 4 & Qcom. X24 LTE & 12.88 & 1.67\\
HTC U12+ & Qcom. X20 LTE & -13.66 & 1.55\\
OnePlus 7T & Qcom. X24 LTE & 12.66 & 1.42\\
iPhone 7 & Intel XMM7360 & -23.86 & 0.88\\
iPhone 8 & Intel XMM 7480& -23.65 & 2.28\\
iPhone X &  Intel XMM7480 & -25.64 & 3.75\\
iPhone 11 & Intel XMM 7660 & -23.19 & 2.49\\
iPhone 11 Pro & Intel XMM 7660 & -25.35 & 2.46 \\\bottomrule
\end{tabular}
\caption{Mobile phones used in the localization and fingerprinting experiments.}
\label{tab:phones}
\end{table}

\begin{figure}[h]
    \centering
    \scalebox{0.67}{\input{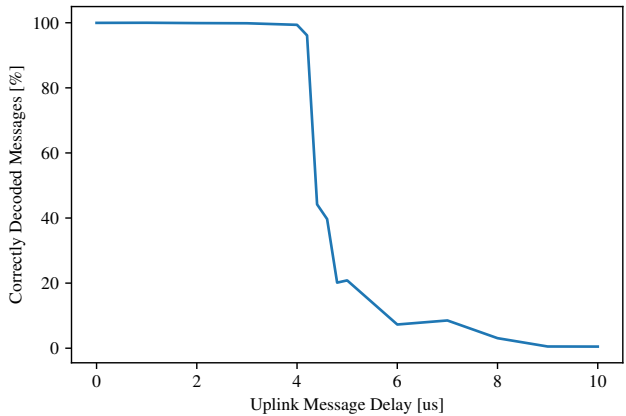}}
    \caption{Percentage of correctly decoded uplink messages by our sniffer as a function of the time delay from the start of a frame.}
    \label{fig:delay}
\end{figure}

%%%%%%%%%%%%%%%%%%%%%%%%%%%%%%%%%%%%%%%%%%%%%%%%%%%%%%%%%%%%%%%%%%%%%%%%%%%%%%%%
\end{document}